\documentclass{article}
\usepackage[utf8]{inputenc}
\usepackage{natbib}
\usepackage{authblk}
\usepackage{float}
\usepackage{titlesec}
\titlelabel{\thetitle.\quad}
\usepackage[colorlinks=true, urlcolor=blue, citecolor=blue, linkcolor=blue]{hyperref}
\usepackage{graphicx}
\graphicspath{ {Figures/} }
\usepackage{caption}
\usepackage{subcaption}
\usepackage{amsmath}
\usepackage{amssymb}
\usepackage{multirow}

\title{An Information Theory Approach to the Stock and Cryptocurrency Market:\\ A Statistical Equilibrium Perspective}
\author[1]{Emanuele Citera}
\author[2,3]{Francesco De Pretis}
\affil[1]{\normalsize Department of Economics, St. Lawrence University, Canton, NY, USA.}
\affil[2]{\normalsize Department of Communication and Economics, University of Modena and Reggio Emilia, Reggio Emilia, Italy.}
\affil[3]{\normalsize Department of Environmental and Occupational Health, School of Public Health, Indiana University Bloomington, Bloomington, IN, USA.}

\date{\today}

\begin{document}
\maketitle

\begin{abstract}
\noindent We study the stochastic structure of cryptocurrency rates of returns as compared to stock returns by focusing on the associated cross-sectional distributions. We build two datasets. The first comprises forty-six major cryptocurrencies, and the second includes all the companies listed in the S\&P 500.
We collect individual data from January 2017 until December 2022. We then apply the Quantal Response Statistical Equilibrium (QRSE) model to recover the cross-sectional frequency distribution of the daily returns of cryptocurrencies and S\&P 500 companies. We study the stochastic structure of these two markets and the properties of investors' behavior over bear and bull trends. Finally, we compare the degree of informational efficiency of these two markets.\\



\noindent \emph{Acknowledgments}: This paper was originally presented at the \textit{Financial Economics Meeting} (FEM 2023) conference held at The EDC Paris Business School, Paris (France), June 29-30, 2023. We wish to thank the participants for their comments and helpful remarks. Errors and omissions are our own.
\end{abstract}

\newpage
\section{Introduction}
Information theory has been widely used to address issues of informational efficiency across different financial markets.

\cite{Batra2022} apply information theoretic measures to study the randomness associations of different financial time series. They study the level of similarities between information-theoretic measures and the various tools of regression analysis, such as the total sum of squares of the dependent variable, relative mutual information and coefficients of correlation, conditional entropy, and residual sum of squares. They observe that mutual information and its dynamic extensions provide an alternative approach with some advantages to studying the association between several international stock indices. Furthermore, they find mutual information and conditional entropy to be relatively efficient compared to the measures of statistical dependence.

\cite{Brouty2022a} introduce a new perspective on investor behaviors in financial markets by using Brillouin's approach to Maxwell's demon \citep{Brillouin1951}. Through a simulation of an agent-based model, they demonstrate that an informed investor using alternative data, which correlates with a financial asset's series of prices, can act as Maxwell's demon in financial markets. This enables them to engage in statistical arbitrage consistently with the Adaptive Market Hypothesis \citep{Lo2019}. In detail, \cite{Brouty2022a} develop a statistical test that assesses the amount of information in a time series using Shannon's Entropy (\citeyear{Shannon1948}). Their main finding is the presence of a recursive cycle between Negentropy (i.e., reverse entropy) and the level of information in the time series in a market with the investor depicted as Maxwell's demon and embedded with the knowledge of alternative data.

%In another contribution, \cite{Brouty2022b} develop a statistical test based on information theory. Their purpose is to test the weak form of market efficiency (which states that current prices reflect all the information contained in past prices) by means of statistical arbitrage. Statistical arbitrage leads to an uncertain output which, on average only, is a gain. In this perspective, they define market efficiency as the absence of statistical arbitrage with predictions based on past prices.

Concerning other information-theoretic approaches, \cite{DelgadoBonal2019} notes that randomness has been quantified in time series using algorithms such as Approximate Entropy (ApEn).\footnote{Although ApEn can be applied to any time series as it is independent of any model, it cannot be used directly to compare financial data series due to the different statistical values observed in the markets.} In their contribution, they extend the use of ApEn to quantify patterns in evolving data series by defining a measure that enables comparisons between time series and epochs using a maximum entropy approach. \cite{DelgadoBonal2019} applies this methodology to the stock market. This demonstrates that the number of patterns changed for the six analyzed markets depending on the economic situation, in agreement with the Adaptive Markets Hypothesis.

\cite{Tran2019} propose a new measure, the Adaptive Market Information Measure (AMIM), to quantify market efficiency. This measure applies to different assets, regions, and data frequencies and allows easy interpretation and comparison. AMIM ranges between zero and one, with a value closer to one indicating lower efficiency and values smaller or equal to zero indicating efficiency. The empirical estimates of AMIM show that markets are often efficient but can also be significantly inefficient over longer periods. The study finds that financial markets become less efficient during major economic events, such as the 2008-09 financial crisis. Overall, \cite{Tran2019} claims that the AMIM measure is simple to compute, robust, and facilitates easy comparisons of efficiency levels across different assets and periods.

Finally, \cite{Vinte2021} apply the intrinsic entropy model as an alternative method to estimate the volatility of stock market indices. Unlike the widely used volatility models that rely solely on a trading day's open, high, low, and close prices (OHLC), the intrinsic entropy model incorporates the traded volumes during the considered time frame. The authors modify the intraday intrinsic entropy model for exchange-traded securities to connect daily OHLC prices with the ratio of the corresponding daily volume to the overall volume traded in the considered period. The intrinsic entropy represents the entropic probability or market credence assigned to the corresponding price level. The intrinsic entropy is calculated using daily historical data for six traded market indices, including S\&P 500, Dow 30, NYSE Composite, NASDAQ Composite, Nikkei 225, and Hang Seng Index. The results obtained by the intrinsic entropy model are compared with the volatility estimates provided by widely used industry estimators. The intrinsic entropy model consistently generates reliable estimates for different time frames and exhibits higher values for the coefficient of variation compared to other advanced volatility estimators. The estimates fall within a significantly lower interval range than those produced by the other models.

The application of information theory has mostly been restricted to traditional financial markets. However, no particular attention has been paid to the cryptocurrency market. In light of these observations, the scope of our paper is twofold. First, by deploying Shannon’s entropy and statistical equilibrium, it attempts to analyze the extent to which the cryptocurrency market can be considered as efficient as the stock market by adopting a cross-sectional perspective. Cross-sectional studies either attempt to explain the price differential of Bitcoin across exchanges located in different countries \citep{Borri2022} or investigate whether conventional cross-sectional characteristics (beta, size, and momentum) explain cryptocurrency returns \citep{Gunther2020}. On the contrary, we focus on the stochastic structure of rates of returns.

Second, the model we develop allows for multiple states, meaning we can assess the extent to which a market can be considered efficient. This is a rather important feature of the model, especially if we compare it to \cite{Brouty2022a}, who develop a two-state model (the market is either efficient or not). Even though the authors acknowledge that including more states makes it more difficult to link entropy to the notion of market efficiency, the two-state choice tends to diminish the power of the statistical test, on the one hand, and lose some information contained in the time series of prices, on the other.

The structure of the paper is the following. Section \ref{data} describes our methodology to collect and analyze the data. Section \ref{model} derives the Quantal Response Statistical Equilibrium model, whereas Section \ref{res} discusses the main results. Finally, Section \ref{concl} concludes and sheds light on future research.

\section{Data Collection and Analysis} \label{data}
To conduct our analysis, we used two samples spanning six years, from January 1, 2017, to December 31, 2022. The samples started in 2017 since many important cryptocurrencies were introduced, and data reliability has significantly increased. 

The first data source is a set of cryptocurrencies, whose details are collected in Appendix \hyperref[cryptos]{A}. We collected the respective daily prices from CoinMarketCap and then computed logarithmic daily returns. We truncate the series of returns at eight standard deviations from the mean. This allows us, on the one hand, to account for potential “black swans”, which are defined as events within three standard deviations from the mean, and, on the other, to focus on the area where the majority of the information lies. Accordingly, we ended up with 98,903 observations. On average, our sample covers 78.17\% of aggregate cryptocurrency market capitalization over the period we considered.

To construct our dataset, we followed \cite{Bhambhwani2023}, who adopted a balanced-panel approach due to a large non-random turnover in the universe of cryptocurrencies.\footnote{Currencies with small capitalization rates disappear as developers abandon the project, miners do not secure their blockchains, or users stop using them. Additionally, many cryptocurrencies are only introduced to capitalize on market upswings through ``pump and dump" schemes. Finally, several currencies have been delisted from exchanges due to fraud or price manipulation.} Our sample comprises forty-six cryptocurrencies, which we can divide into three groups: 11 Proof-of-Work, 7 Non-Proof-of-Work, and 28 additional. Following \cite{Bhambhwani2023}, we selected the first two groups from CoinMetrics, which only reports price data from the most reputable cryptocurrency exchanges and uses thirty-five criteria to filter out illiquid or unreliable exchanges. Instead, the remaining 28 cryptocurrencies were selected from Bittrex, listed as one of the trusted exchanges by Bitwise in their report to the SEC regarding inflated and wash trading volumes on exchanges.

\begin{figure}[H]
    \centering
    \caption{Pooled Distributions.}
    \includegraphics[width=.75\textwidth]{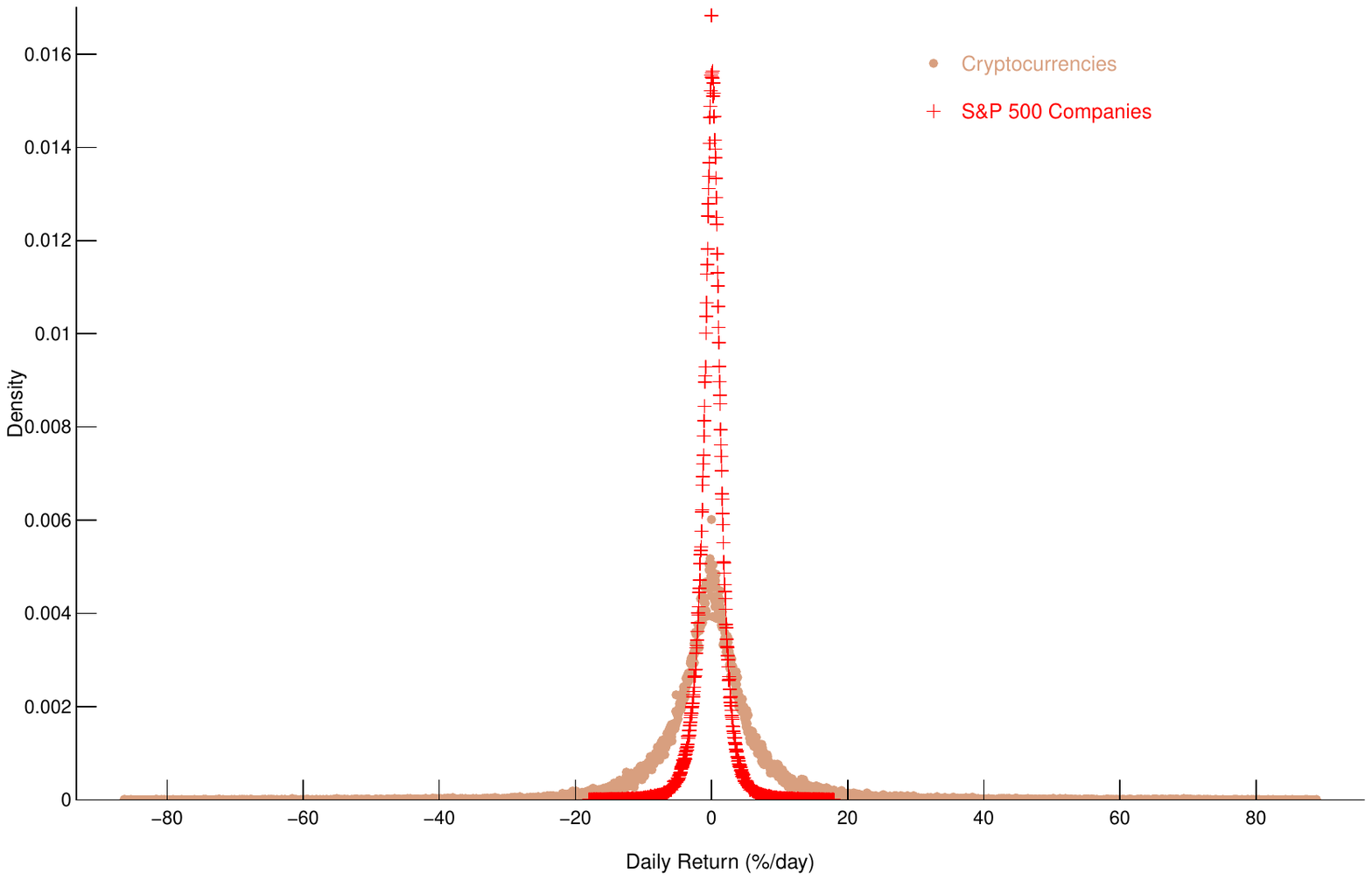}
    \\ \small Source: CoinMarketCap and Yahoo Finance.
    \label{fig:pd}
\end{figure}

The second data set collects daily closing adjusted prices of the companies listed in the S\&P 500, which we retrieved from Yahoo Finance. We compute logarithmic daily returns at the individual company level and then truncate the series at eight standard deviations from the mean. Our final dataset is made of 748,017 observations, which corresponds to 99.87\% of the original sample. Figure \ref{fig:pd} compares the pooled distributions of crypto and S\&P 500 companies' daily returns. Table \ref{sum_pd} summarizes the main summary statistics and the four moments of the distributions.

Despite the average return of the two classes of assets is about the same, the range of variation of cryptocurrencies is as five times as bigger than that of individual companies. This is not surprising, given that the cryptocurrency market's volatility is considerably higher than the stock market. We can also acknowledge significant differences in the second and fourth moments of the distributions. In particular, whereas S\&P 500 companies' returns are negatively skewed, this is not the case for cryptocurrencies. Furthermore, our cryptocurrency dataset exhibits significantly higher kurtosis than equities' returns. We attribute the discrepancy to the structural features of the cryptocurrency market, which is subject to less regulation and an increased arbitrage across exchanges \citep{MakarovSchoar2020}.

\begin{figure}[H]
    \centering
    \caption{Crypto Vs S\&P 500 Index. (Red bars denote bear markets).}
    \begin{subfigure}[b]{\textwidth}
        \centering
        \includegraphics[width=.7\textwidth]{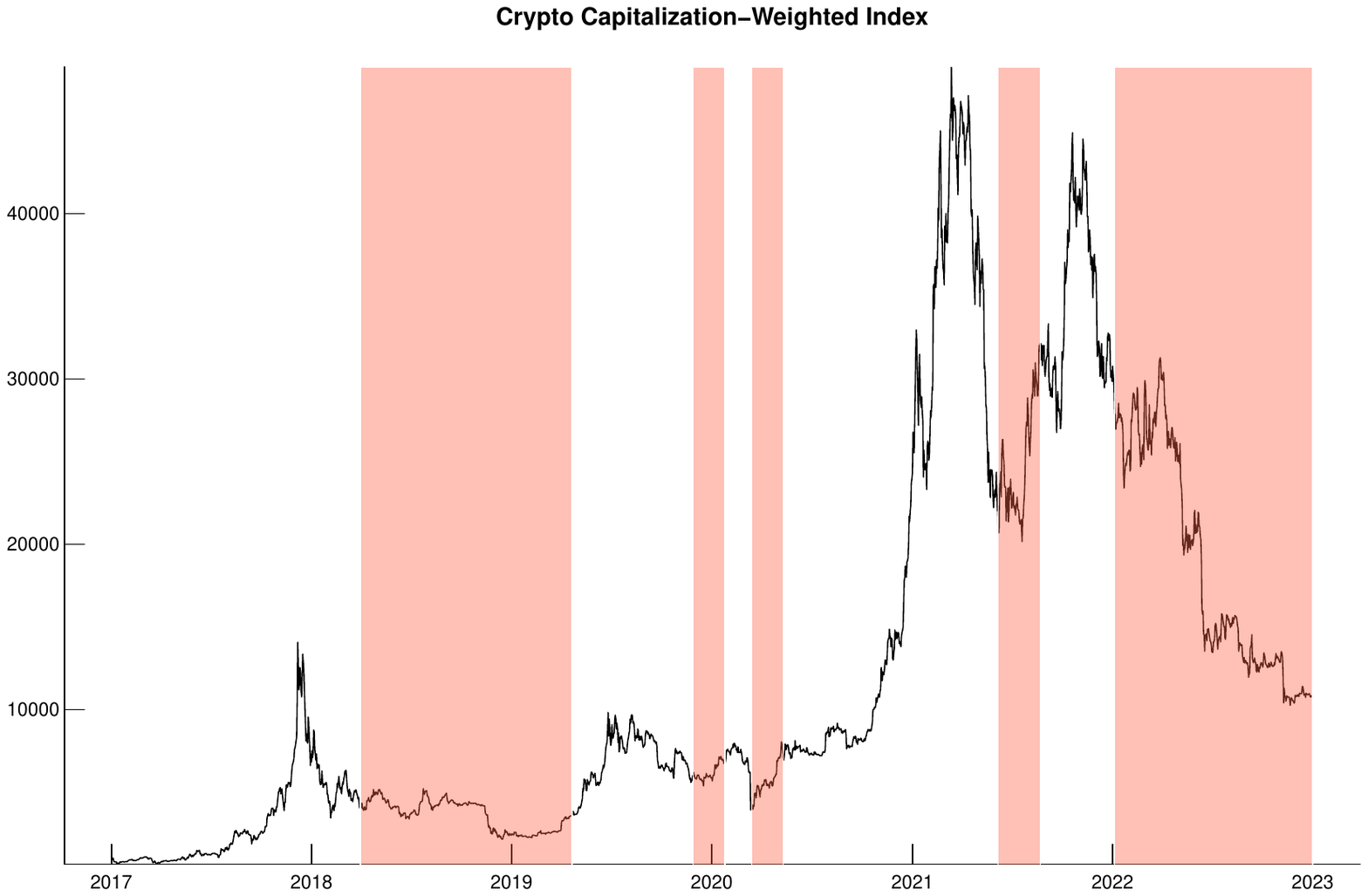}
     \end{subfigure}
     \begin{subfigure}[b]{\textwidth}
        \centering
        \includegraphics[width=.7\textwidth]{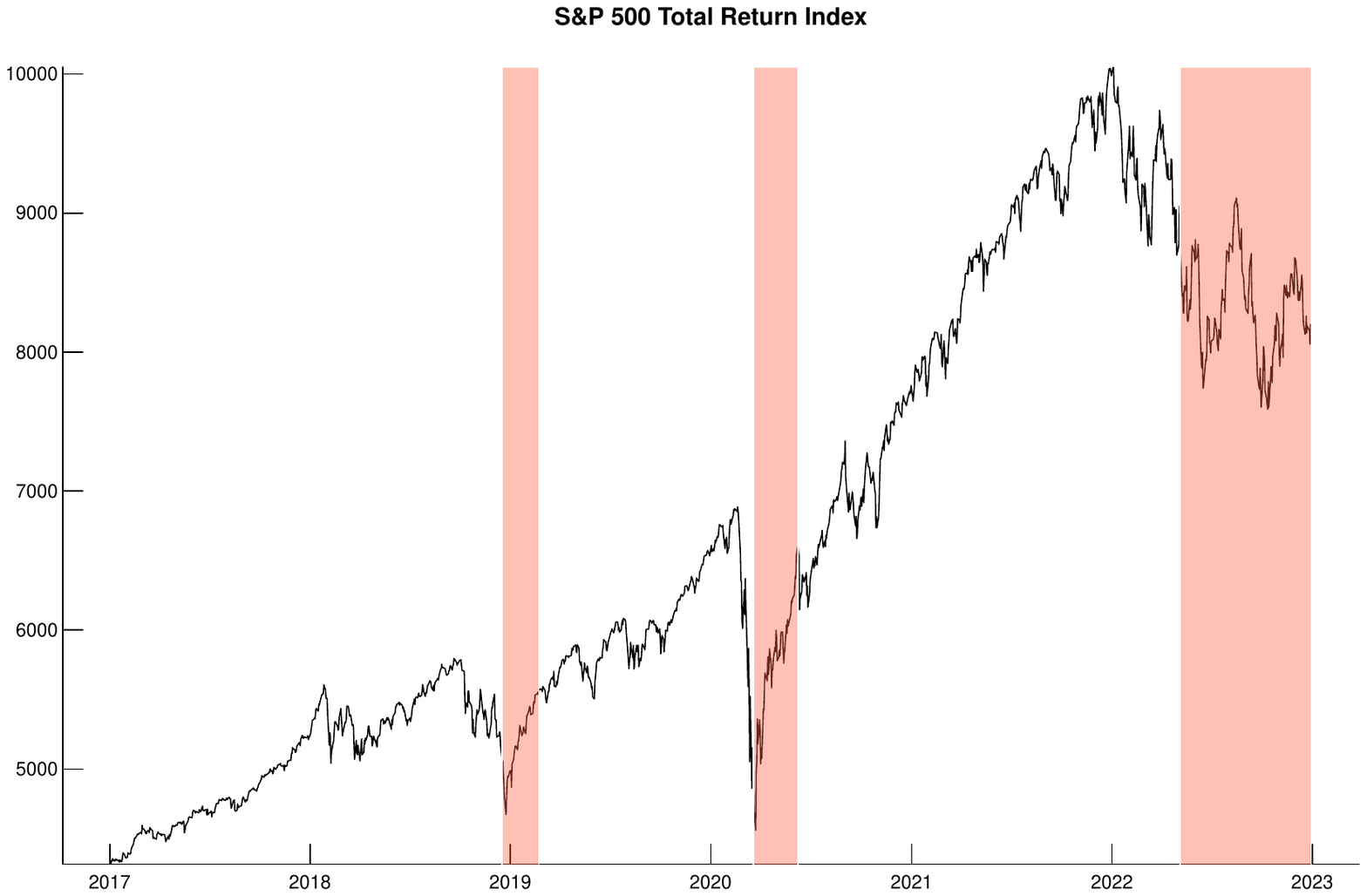}
     \end{subfigure}
    \\ \small Source: CoinMarketCap and Yahoo Finance.
    \label{fig:index_comp}
\end{figure}

To better understand the stochastic behavior of the two classes of financial assets, we identify the associated bull and bear market trends. We define bear and bull markets according to the 50-day and 200-day exponential moving average (EMA) crossover.\footnote{Given that we are comparing two markets with completely different scales of fluctuations, we do not follow the standard definition provided by the Securities and Exchange Commission, which defines bear markets as declines of 20\% or more over at least two months.} In detail, we have a bull market when the 50-day EMA lies above the 200-day, and a bear market in the opposite scenario. Subsequently, we construct a cryptocurrency capitalization-weighted index with daily closing prices and market capitalization of individual crypto assets. As per the dataset concerning the companies listed in the S\&P 500, we use the closing price of the S\&P 500 Total Return Index. Figure \ref{fig:index_comp} shows the resulting trends in both series. Whereas the S\&P 500 records few instances of bear markets, the cryptocurrency market generates more occurrences of bear trends over prolonged periods.

\begin{table}[H]
    \centering
    \caption{Summary Statistics of Pooled Distributions (\%/day).}
    \begin{tabular}{cccccccc}
    \hline \\[-1.8ex]
    & Min & Mean & Median & StDev & Skew & Kurt & Max\\ [.5ex]
    \hline  \hline \\[-1ex]
    Cryptocurrencies & -86.33 & 0.04 & -0.09 & 8.56 & 0.58 & 13.27 & 88.87\\
    S\&P 500 Companies & -17.99 & 0.05 & 0.09 & 2.11 & -0.14 & 8.41 & 17.98\\
    [1ex] \hline
  \end{tabular}
  \label{sum_pd}
\end{table}

\subsection{Stationarity and Reversibility of time-series}
To assess whether our two datasets could be assumed to be under statistical equilibrium, we checked two different types of statistical properties: (a) if the time series were stationary or trend stationary for given time windows -- a standard approach usually employed in econometrics; (b) if they were an expression of time-reversible processes. For the latter, we follow \cite{Kardar2007}, who points out that a reversible process can be run backward in time by simply reversing its inputs and outputs. Such a process is the thermodynamic equivalent of a frictionless motion in mechanics. Since time reversibility implies equilibrium, a reversible transformation must be quasi-static, even if the reverse is not necessarily true. The next subsections briefly outline some tests we used to check for such properties.

\subsubsection{Stationarity and trend-stationarity tests}
The Augmented Dickey-Fuller (ADF) and Kwiatkowski-Phillips-Schmidt-Shin (KPSS) tests are commonly used to determine the stationarity or trend-stationarity of a time series. We detail them below and provide information about the results obtained over our empirical data.\\

\noindent \textbf{Augmented Dickey Fuller (ADF) Test}: The ADF test \citep{Dickey1979,Dickey1981} is a type of unit root test. A unit root test determines how strongly a trend defines a time series. The ADF test has the null hypothesis that a unit root is present in a time series sample. If the null hypothesis of the ADF test cannot be rejected, this suggests that the time series has a unit root, implying that it is non-stationary. The ``augmented" aspect of the test includes lagged terms of the time series to account for serial correlation. In this test, the Null Hypothesis (H0) reads that the time series has a unit root (i.e., non-stationary). In contrast, the Alternative Hypothesis (H1) states that the time series does not have a unit root (i.e., stationary). If the test statistic is less than the critical value, we reject the null hypothesis, suggesting the time series is stationary.\\

\noindent \textbf{Kwiatkowski-Phillips-Schmidt-Shin (KPSS) Test}: The KPSS test \citep{Kwiatkowski1992} is another method to test for stationarity, but its hypotheses are framed differently from the ADF test. The KPSS test has the null hypothesis that a time series is trend-stationary around a deterministic trend. This means that the time series can be made stationary by detrending it (removing the deterministic trend). In this test, the Null Hypothesis (H0) states that the time series is stationary around a deterministic trend. In contrast, the Alternative Hypothesis (H1) reads that the time series has a unit root (i.e., non-stationary). If the test statistic is greater than the critical value, we reject the null hypothesis, suggesting the time series is non-stationary.\\

\noindent Often, both ADF and KPSS tests are used together. The reason is that they complement each other. ADF focuses on the null being a unit root (non-stationary), while KPSS assumes stationarity as the null. If both tests indicate stationarity, then the series is stationary. Conversely, if both tests indicate non-stationarity, then the series is non-stationary. Eventually, if the results conflict, further diagnostics or considerations might be necessary.\\

\begin{figure}[H]
    \centering
    \caption{Stationarity Tests for the S\&P 500 Total Return Index.}
    \includegraphics[width=\textwidth]{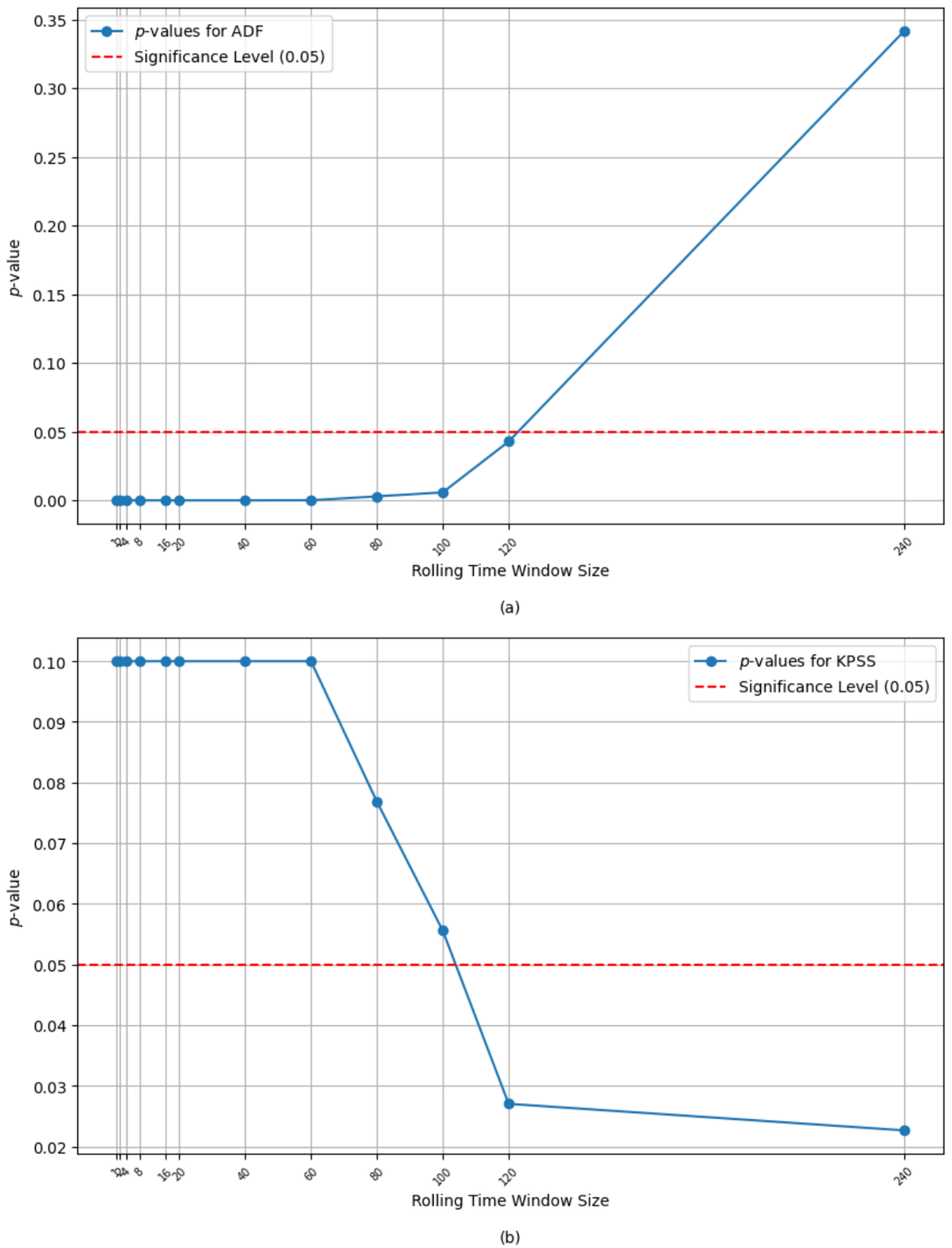}
    \\ \small $p$-values for stationarity tests over different rolling time windows.\\Panel (a): ADF stationarity test. Panel (b): KPSS stationarity test. 
    \label{fig:st}
\end{figure}

\noindent Figure \ref{fig:st} outlines the results of those tests for the S\&P 500 index, where data appears to be stationary and trend-stationarity up to five trading months (100 days rolling windows). For the cryptocurrency index, we did not find comparable results in terms of stationarity, probably due to the very nature of that market, which has different trading volumes and is less mature than the stock market. Using the latter's results as our benchmark, we now perform time-reversible tests.

\subsubsection{Irreversibility tests}
Several irreversibility tests are known in the literature. \cite{Zanin2021} has provided an extensive review. We utilize four of the most commonly used tests for our analysis, detailing their methodology and presenting the results derived from our data.\\

\noindent \textbf{Ramsey and Rothman's Test}: Ramsey and Rothman's irreversibility test \citep{Ramsey1996} is a statistical procedure used to detect the presence of asymmetric adjustments towards long-run equilibrium in economic variables. This test specifically checks whether positive and negative shocks to a system have different effects on its return to equilibrium, implying that the system does not symmetrically revert to the mean. In other words, the test determines where a system reacts differently when above or below equilibrium, shedding light on the nature of adjustments and potential policy implications in various economic contexts. Note that this test employs one degree of freedom, with $k$ identifying different time lags.\\

\noindent \textbf{Daw, Finney, and Kennel (DFK) test}: The DFK test \citep{Daw2000} is a symbolic method designed to measure temporal ``irreversibility" in time-series data. Temporal irreversibility indicates that the dynamics of a system are not merely Gaussian linear or a static transformation thereof. The DFK test employs an algorithm named ``false flipped symbols" to detect this irreversibility. This method is favored because of its computational efficiency, noise resilience, and statistical analysis ease. The test can help characterize experimental data and is particularly useful for choosing alternative models. We employ two degrees of freedom in our analysis: $n$, the number of possible symbols, and $L$, the word length.\\

\noindent \textbf{Enhanced Visibility Graphs}: The enhanced Visibility Graphs \citep{Luque2009}, specifically the horizontal visibility algorithm as in the version improved by \cite{Zanin2021}, is a method that transforms time series data into complex networks. By doing so, one can employ complex network theory techniques to characterize and understand the nature of the time series. Some key points about this approach:

\noindent \textit{Mapping of Random Series}: The horizontal visibility algorithm primarily deals with random series (series made of independent identically distributed random variables). The algorithm's properties provide detailed insights into the topological attributes of graphs associated with these random series, such as the degree distribution, clustering coefficient, and mean path length.

\noindent \textit{Discriminating Randomness}: A distinct feature of this method is its ability to distinguish randomness in time series. Specifically, any random series maps onto a graph where the exponential degree distribution follows \(P(k) = \frac{1}{3} (\frac{2}{3})^{k-2}\). This distribution remains consistent regardless of the original probability distribution of the series. Hence, any visibility graph with a different \(P(k)\) indicates the presence of non-randomness in the series.

\noindent \textit{Distinguishing Chaotic Series}: Beyond just random series, the horizontal visibility algorithm can differentiate between chaotic series and those based on the independent and identically distributed (i.i.d.) theory. The algorithm is effective in analyzing (i) noise-free low-dimensional chaotic series, (ii) noisy low-dimensional chaotic series, even with significant noise, and (iii) high-dimensional chaotic series, all without requiring additional methods or noise reduction techniques.

\noindent \textit{Analysis of Conjectured Sequences}: This method also offers heuristic explanations for the topological properties of chaotic series and can analyze random series.\\

\noindent \textbf{Local Clustering Coefficient}: The Local Clustering Coefficient test \citep{Donges2013} is designed to detect time-reversal asymmetry in nonlinear time series. This asymmetry is a defining characteristic of many nonlinear series, indicating that the series does not simply reverse when time runs backward. Key aspects of this method are:

\noindent \textit{Basis on Visibility Graphs}: The test employs standard and horizontal visibility graphs as foundational tools (as outlined above). These graphs transform time series data into complex network representations, allowing analysis using network theory measures.

\noindent \textit{Time-Directed Network Measures}: Central to this method is the comparison of the distributions of time-directed variants of standard complex network measures, specifically the degree and local clustering coefficient. When adjusted to consider the direction of time, these measures can reveal asymmetry or irreversibility in the series.

\noindent \textit{No Need for Surrogate Data}: One advantage of this test is that it does not rely on surrogate data. Surrogate data techniques often involve generating randomized versions of the original time series to test against, which can be computationally expensive and may not always provide clear contrast.

\noindent \textit{Applicability to Short Series}: This approach can be effectively applied even to relatively short time series, making it versatile for different datasets.\\

\begin{figure}[H]
    \centering
    \caption{Irreversibility tests for cryptocurrency index and S\&P 500 Total Return Index.}
    \includegraphics[width=1.3\textwidth]{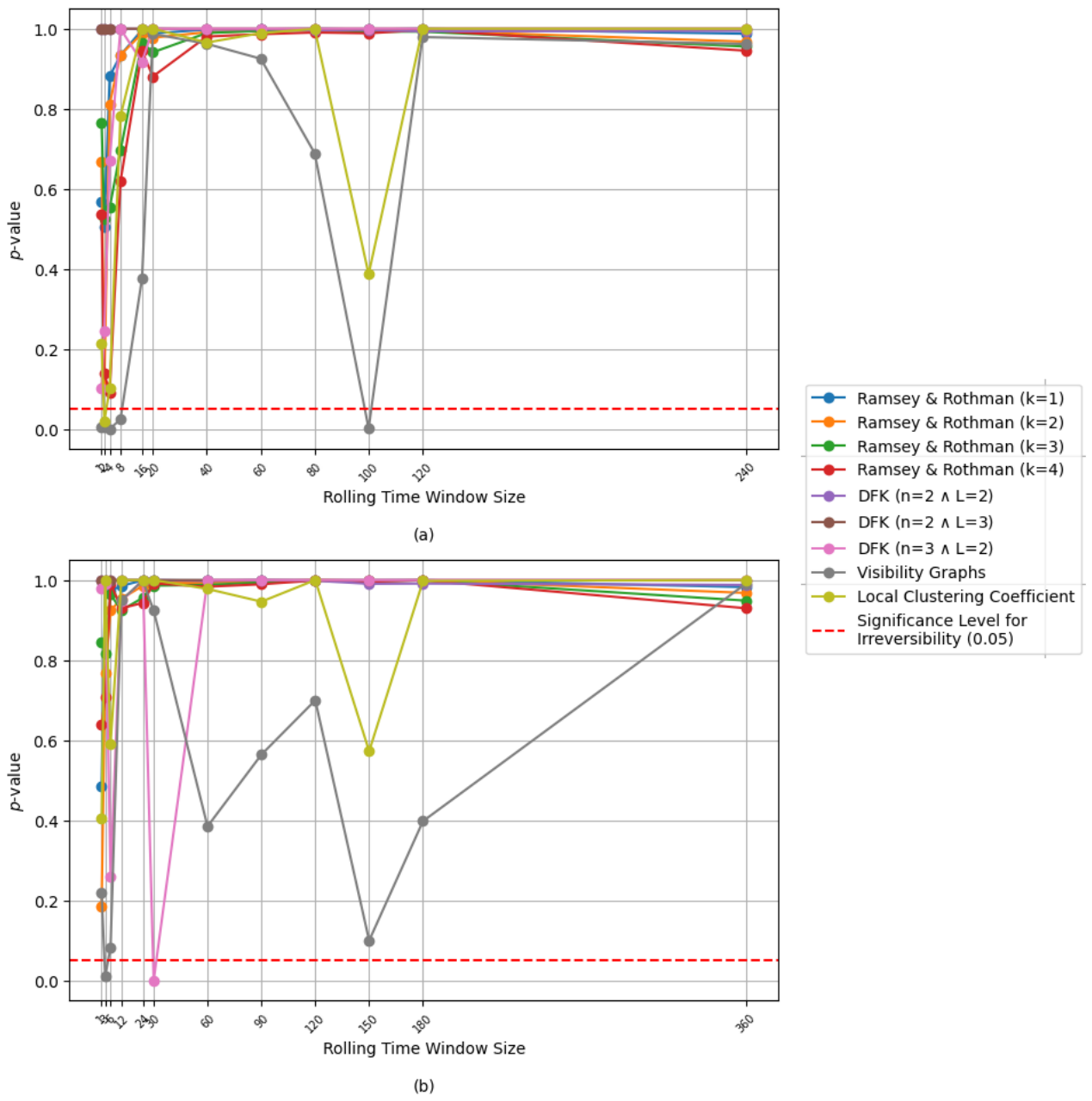}
    \\ \small $p$-values for irreversibility tests over different rolling time windows.\\Panel (a): S\&P 500 Total Return Index. Panel (b): Cryptocurrency Index. 
    \label{fig:it}
\end{figure}

\noindent Figure \ref{fig:it} shows the results for those tests as applied to our cryptocurrency index and the S\&P 500 Total Return Index. As we can see, within the previously identified window of five trading months, the four trading month window (80 days for the S\&P data and 120 days for cryptocurrencies) seems to be a good candidate for our analysis. Indeed, it provides a balanced horizon between reversible and irreversible areas, i.e., above and below the significance level of the tests (red line). Accordingly, we select this time window for our analysis in the following sections.

\newpage
\section{The Quantal Response Statistical Equilibrium Model} \label{model}
The Quantal Response Statistical Equilibrium (QRSE) model, originally put forward by \cite{SF2017}, provides a method to explain the observed regularities in highly complex systems (see also \cite{Zheng2023}) by taking into account social interactions between a large number of heterogeneous individuals and their reactions to the economic variable of interest (in our case, the rate of return on asset prices). The main objective is to analyze the market dynamics without relying on ad-hoc assumptions.

In the QRSE model, equilibrium takes the form of an information-theoretic probability distribution representing all possible states of the system. By adopting the Maximum Entropy Principle \citep{Jaynes2003} as a method of inference, which we discuss in more detail in Appendix \hyperref[MEP]{B}, the model introduces a general framework where the conventional general equilibrium outcome exists just as a special and highly unlikely state of the economic systems.%\footnote{An interactive version of the model is available here: \href{https://emanuelecitera.shinyapps.io/Shiny/}{https://emanuelecitera.shinyapps.io/Shiny/}}.
The underlying logic of the model is based on classical political economic theory \citep{Smith1776}, and we apply it to the rate of return on asset prices as follows.

Investors seeking above-average returns from their transactions generate a ``tendential gravitation" around a certain rate as an unintentional result of their interactions with other actors. Therefore, their decisions to compete for higher expected returns determine, at any point in time, the change in price level, thus creating an average rate of price change in the market as a whole. This process, which can be interpreted as the statistical equalization of returns, generates statistical regularities observed in the fat-tailed, single-peaked distributions characterizing stock returns over different time horizons.

If we relate our model's logic to the behavior of financial markets, we can identify three components of crucial importance. First, investors' alertness to excess returns is an important determinant of the market's volatility. Second, the unintended feedback of transactions on returns dictates how trading impacts liquidity, affecting variations in the return level. Third, investors' expectations determine whether transactions occur, thus relating individual beliefs to market outcomes. The QRSE model can help explain the above features by accounting for (1) the assumption of quantal response behavior of the market participants, (2) the negative feedback mechanism (representing the impact of individual actions on social outcomes), (3) the role of expectations. Let us now turn to analyze each concept in more detail.

\subsection{The Behavioral Constraint}
\begin{figure}[H]
    \centering
    \caption{Logit Quantal Response Conditional Probabilities.}
	\includegraphics[width=.8\linewidth]{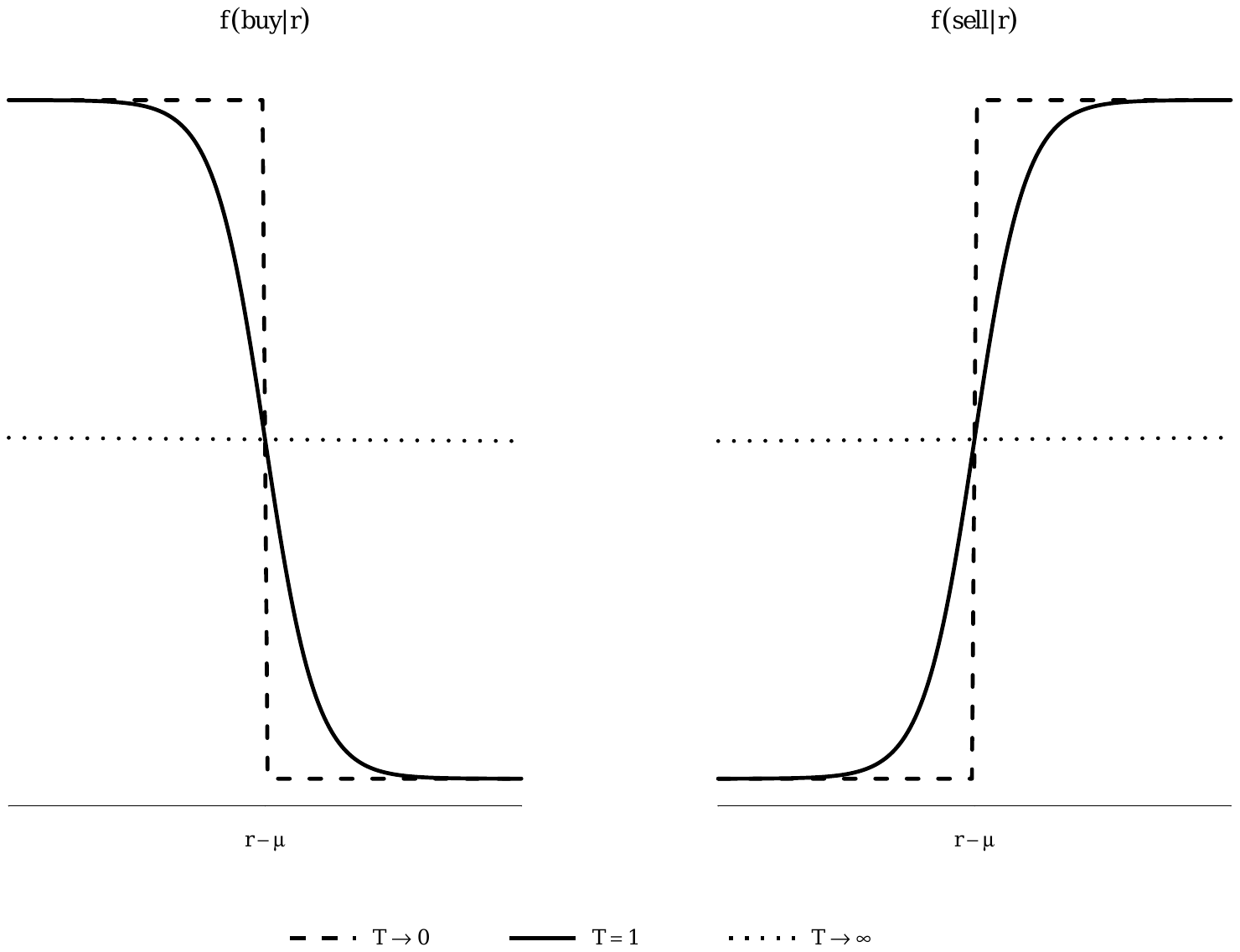}
	\\ \footnotesize The functions are plotted for different values of $T$, while holding $\mu=0$.
	\label{fig:bc}
\end{figure}

The first component of the model represents a behavioral theory of the typical agent in terms of the probability that they will choose a particular action, $a=\{sell,buy\}$, conditional  on the variable $r$, the rate of return, and is expressed in the conditional distribution $f[a|r]$. This conditional distribution expresses the response probabilities over the action set given $r$, and quantifies the impact of $r$ on the individual action frequencies. Its functional form, whose analytical derivation is shown in Appendix \hyperref[bc]{C}, is the logit quantal response distribution (with $\beta=\frac{1}{T}$):
\begin{align} \label{eq:lqr}
    f[buy|r]&=\frac{1}{1+e^{\frac{u[a,r]}{T}}} &
    f[sell|r]&=\frac{1}{1+e^{-\frac{u[a,r]}{T}}}
\end{align}
The parameter $T$ represents the attentiveness of the typical agent. As Figure \ref{fig:bc} shows, the lower the $T$, the more alert the individual is to differences in payoff, and the more closely the action approximates the unconstrained payoff-maximizing outcome.

\begin{figure}[H]
	\centering
	\caption{Relationship between $T$ and QRSE Distribution ($\hat{f}[r]$).}
	\begin{subfigure}[b]{\textwidth}
        \centering
		\includegraphics[width=.65\textwidth]{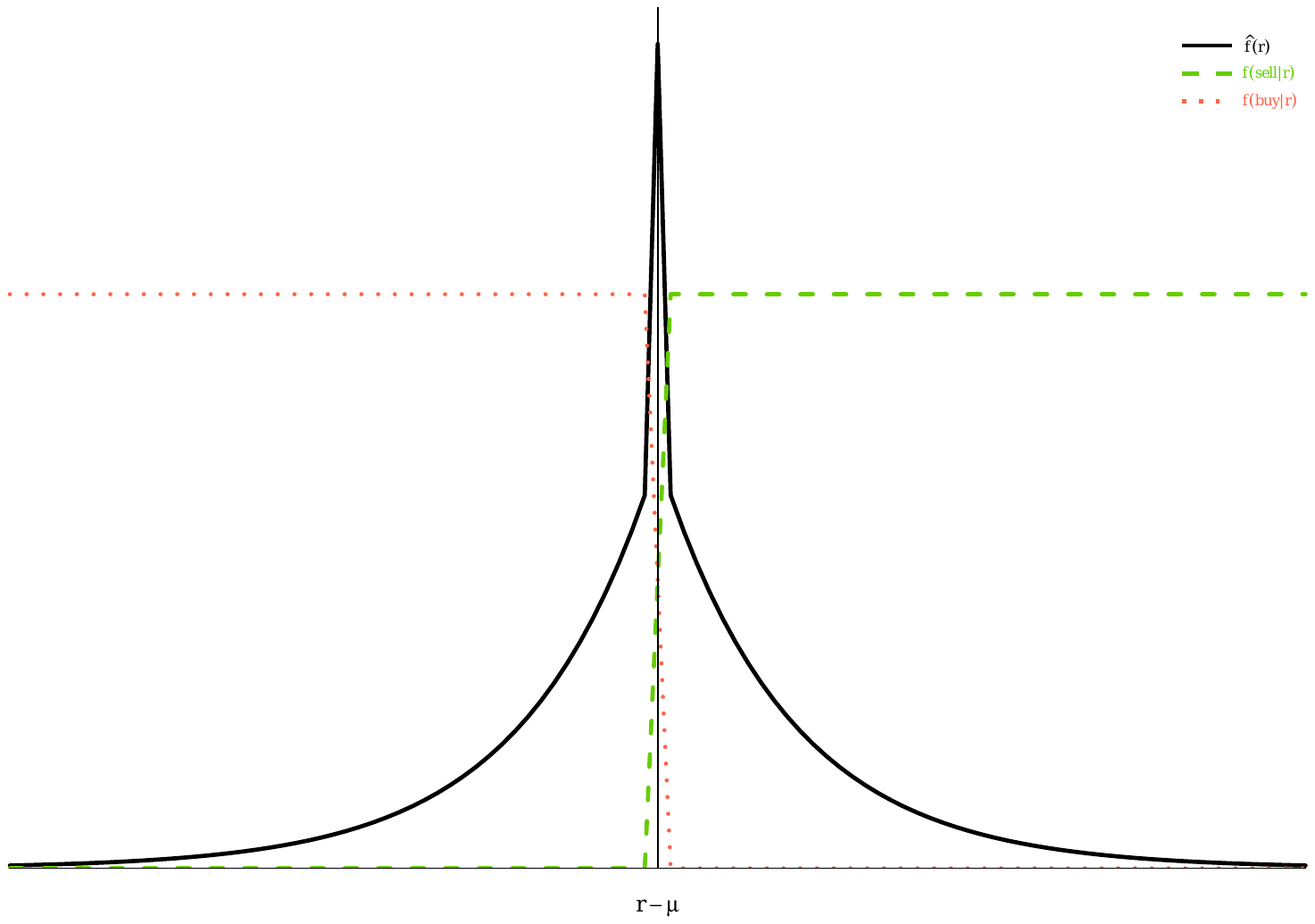}
		\caption{\scriptsize $\mu=0,\, T=0.0001$.}
		\label{fig:lqr_0}
	\end{subfigure}
	\begin{subfigure}[b]{\textwidth}
        \centering
		\includegraphics[width=.65\textwidth]{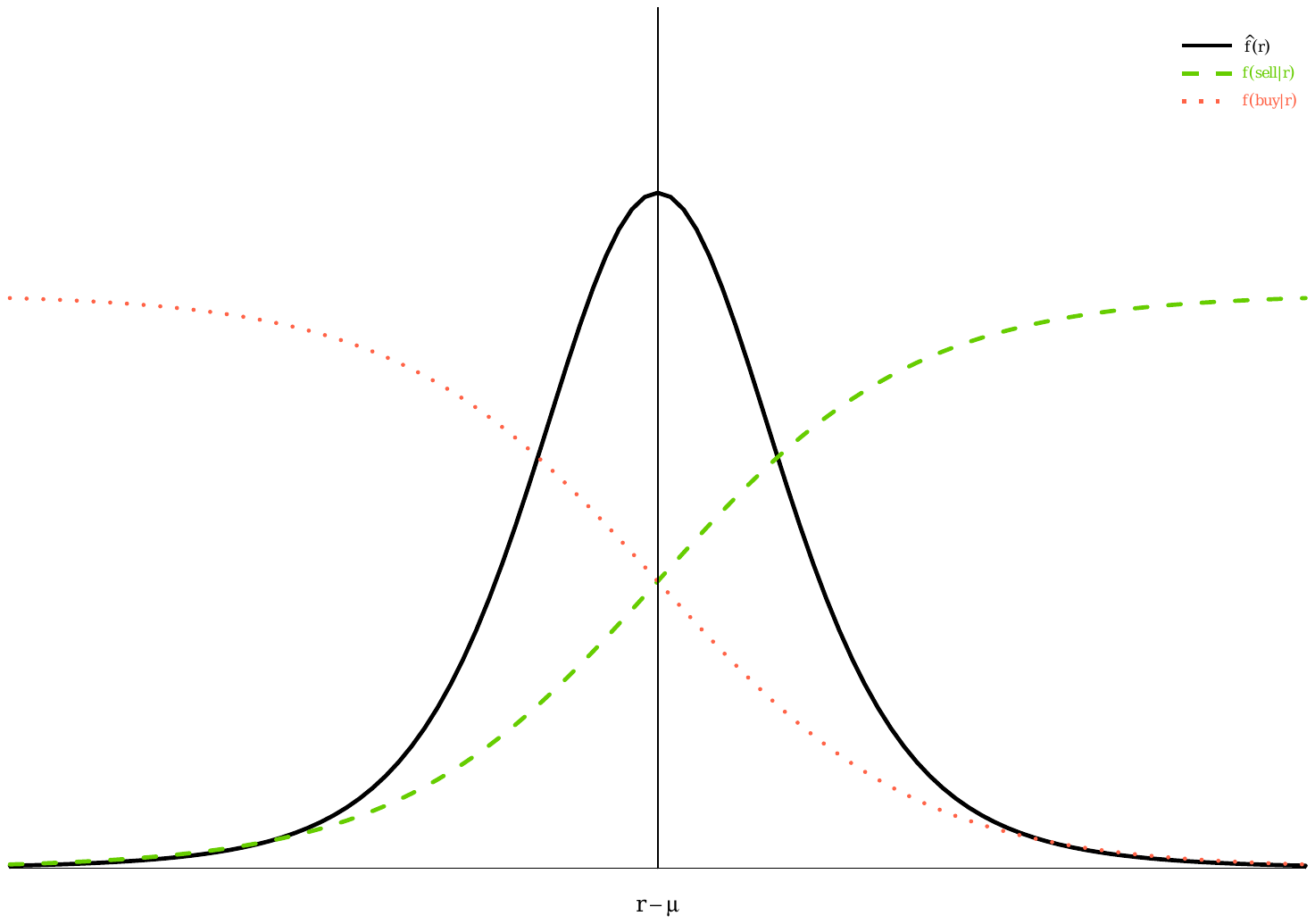}
		\caption{\scriptsize $\mu=0,\, T=1$.}
		\label{fig:lqr_1}
	\end{subfigure}
	\begin{subfigure}[b]{\textwidth}
        \centering
		\includegraphics[width=.65\textwidth]{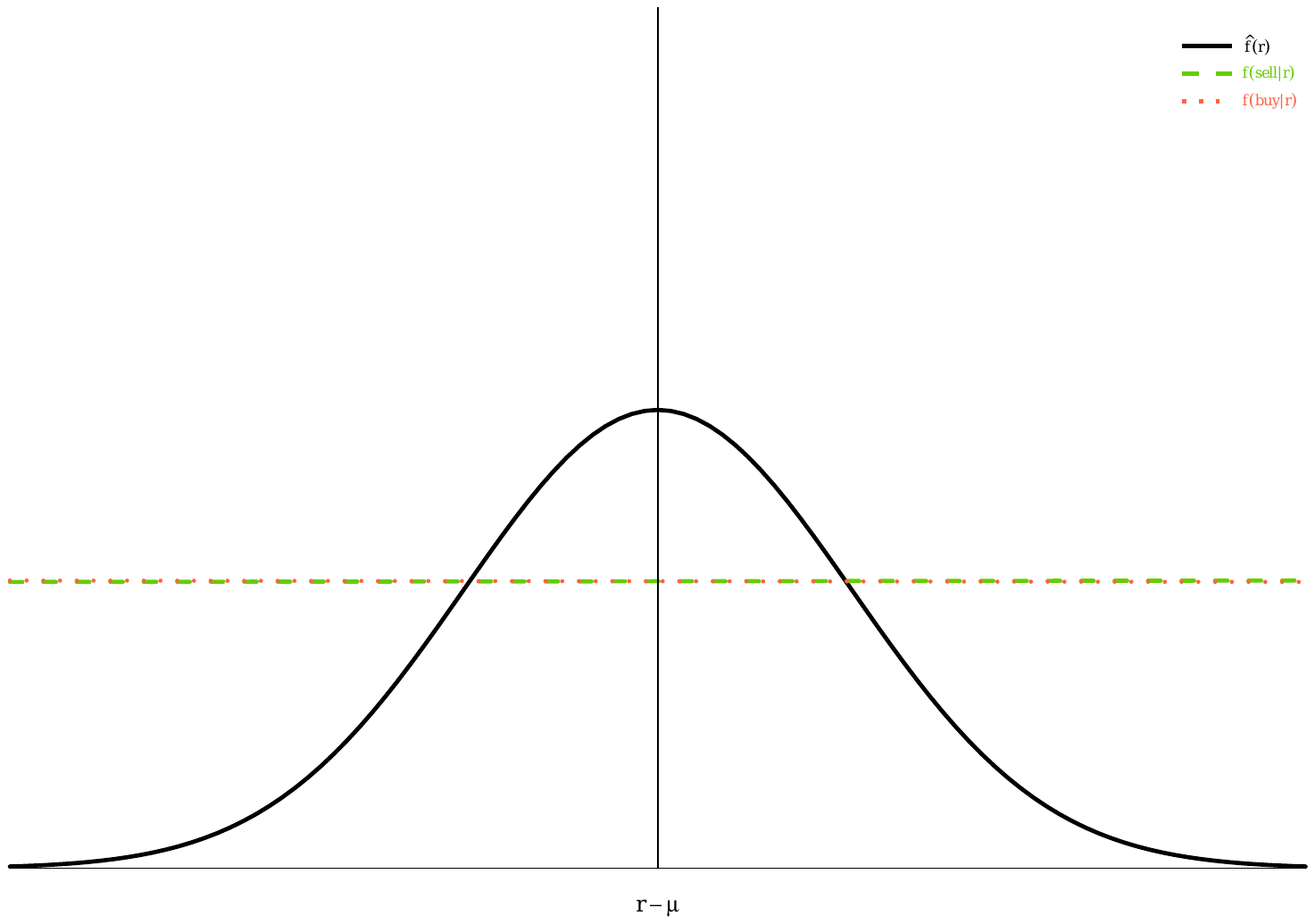}
		\caption{\scriptsize $\mu=0,\, T=1000$.}
		\label{fig:lqr_max}
	\end{subfigure}
	\label{fig:lqr}
\end{figure}

While entropy-constrained behavior is a generalization of rational choice theory, there are several significant differences with crucial implications \citep{Faynzilberg1996}. We show them in Figure \ref{fig:lqr}, and explain them in what follows.

In the limiting case where $T\to0$ (Figure \ref{fig:lqr_0}), the behavior of the agent becomes completely deterministic, meaning that agents will always choose to sell an asset whenever $r>\mu$ and to buy it whenever $r<\mu$. In turn, the conditional distribution $f[a|r]$ approaches the Heavyside step function, and the QRSE distribution ($\hat{f}[r])$ approximates a single-peaked Laplace distribution.

When $T>0$, individual agents' preferences no longer satisfy the assumptions of consistency and completeness, and there is a positive probability for each action an individual will make. In this case, $\hat{f}[r]$ will exhibit fat-tails similar to a Student’s T distribution (Figure \ref{fig:lqr_1}).

Finally, as $T\to\infty$ the actions are random and conditionally independent of the outcome $r$, and therefore $f[a|r] \to f[a]$. As a result, the QRSE marginal distribution $\hat{f}[r]$ converges to a Normal distribution, as shown in Figure \ref{fig:lqr_max}.%\footnote{Note that, in these three cases, we are assuming a positive impact of the actions of buying/selling on stock returns. Indeed, as we will see later, the standard deviation of $\hat{f}[r]$ is also regulated by another parameter, $S$, that here we hold constant and equal to 1}.

In terms of investors' behavior, the above results have two implications. First, they dovetail with some behavioral theories, such as the Adaptive Market Hypothesis \citep{Lo2019}, which calls for adaptive and evolutionary investors' behavior as market conditions change. In our framework, this can be captured by changes in the degree of attentiveness, which allows for heterogeneous behaviors. Second, a characteristic of logistic behavior models is that the log odds of making two decisions are equal to the ratio of the expected utility of the two decisions, which implies:
\begin{equation} \label{eq:log_odds}
\log\left[\frac{f[buy|r]}{f[sell|r]}\right]= \frac{r-\mu}{T} \, .
\end{equation}

The above equation tells us something about the risk propensity of the agent. Indeed, since the parameter $T$ denotes the standard deviation of the Sigmoid function (Equation \ref{eq:lqr}), the higher the $T$, the higher the risk a trader is willing to accept in making investment decisions, and vice versa.

\subsection{The Feedback Constraint}
Economic variables that exhibit statistical regularities, such as a central moment and stable endogenous fluctuations, can be considered in equilibrium precisely because there are negative forces that stabilize the distribution. Indeed, whereas payoff maximizing agents are seeking rates of return above their expected fundamental valuation of the asset to sell, buyers look for stocks whose returns are below their expected fundamental rate of price increase. Unintentionally, sellers' actions cause a decline in stock returns, while buyers' behavior causes an increase. Accordingly, the second component of the model reflects the impact of the action on the outcome variable, $f[r|a]$.

The conditional distribution $f[r|a]$ expresses a theory of the formation of social outcomes. If the actions of individual participants had no impact on the outcome, then $f[r|a] = f[a]$, as in the standard general equilibrium framework. However, in light of our previous discussion, we assume that actions impact outcomes, which implies that buying/selling a stock tends to increase/decrease its rate of return.

Given that statistical equilibrium in the joint distribution, $f[a,r]$, implies that $r$ is statistically regulated by $a$ through a negative feedback mechanism, the buying and selling actions tend to push the market outcomes around a certain level of return, which we denote with $\alpha$. Accordingly, we can write this condition as follows:
\begin{equation}
\int f[buy,r]r \,dr \leq \alpha \leq \int f[sell,r]r \,dr\\
\end{equation}

Then, we have:
\begin{align}
\begin{split}
&\int f[sell,r](r-\alpha) \,dr - \int f[buy,r](r-\alpha) \,dr\\
=&f[sell]\mathbb{E}[(r-\alpha)|sell] - f[buy]\mathbb{E}[(r-\alpha)|buy]\\
=&\int (f[sell|r] - f[buy|r])f[r](r-\alpha) \,dr\\
=&\int \left(\frac{1}{1+e^{-\frac{u[a,r]}{T}}} - \frac{1}{1+e^{\frac{u[a,r]}{T}}}\right) f[r](r-\alpha) \, dr\\
=& \int \left(\frac{1-e^{-\frac{u[a,r]}{T}}}{1+e^{-\frac{u[a,r]}{T}}}\right) f[r](r-\alpha) \, dr\\
=&\int \tanh \left[\frac{r-\mu}{2T}\right] f[r](r-\alpha) \,dr \geq 0
\end{split}
\end{align}

Note that if there were no impact of the actions on the outcome, the expectation of stock returns conditional on the buying action would tend to be higher than the expectation conditional on the selling action. However, the presence of negative feedback tends to constrain this difference to a positive but finite value $\delta$, implying that:
\begin{equation} \label{eq:fc}
\int f[sell,r](r-\alpha) \,dr - \int f[buy,r](r-\alpha) \,dr \leq \delta
\end{equation}

\begin{figure}[H]
    \centering
	\caption{The negative feedback mechanism.}
	\includegraphics[width=.8\linewidth]{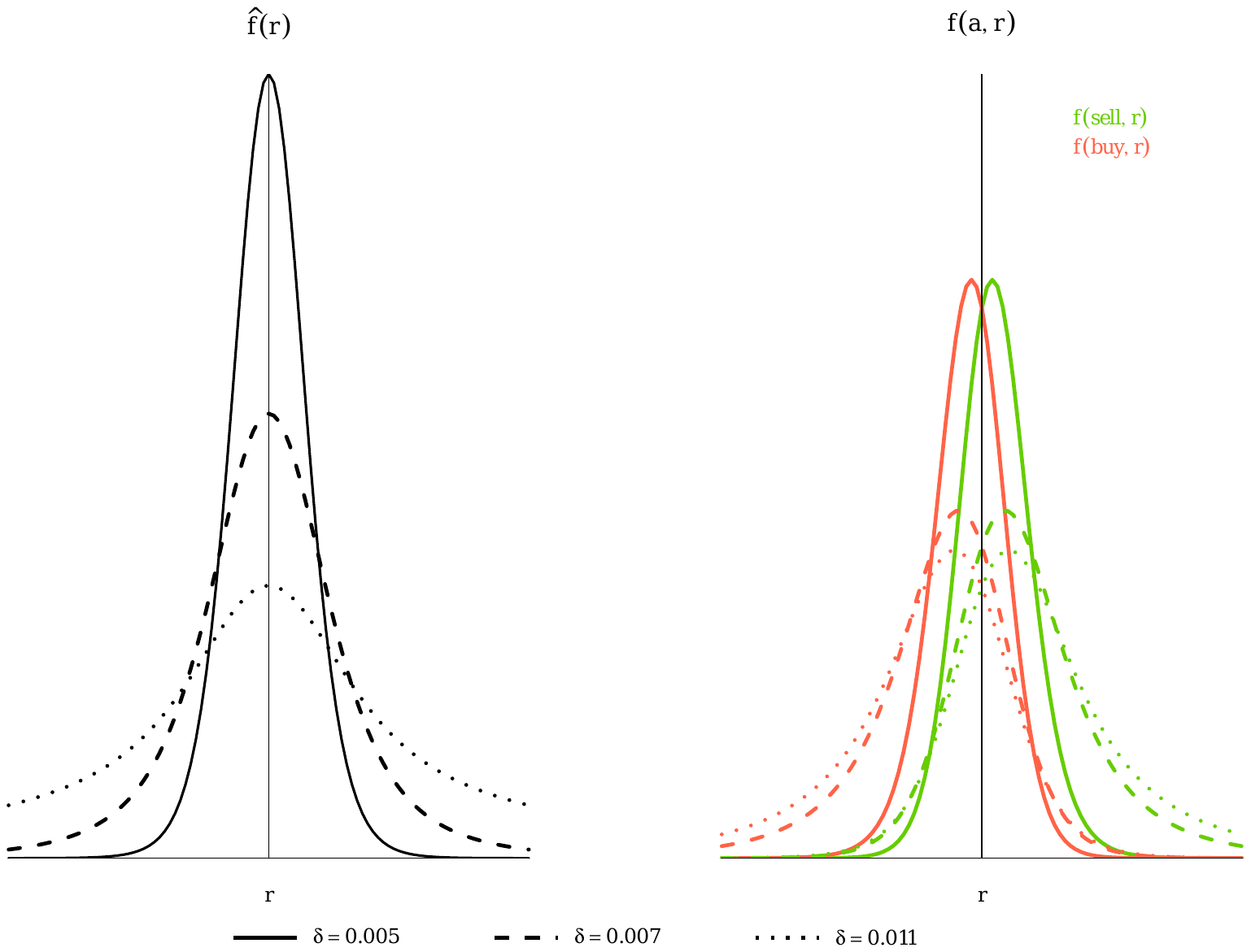}
	\label{fig:nf}
\end{figure}

The parameter $\delta$ is an indirect measure of the dependence of $a$ on $r$. The smaller the $\delta$, the stronger the actions in changing the outcome. When $\delta=0$, we have:
\begin{equation*}
    \int f[sell,r]r \,dr = \int f[buy,r]r \,dr =\alpha \, .
\end{equation*}
In this case, the actions have an infinite effect on stabilizing the outcome at the $\alpha$ level of stock return. As a result, the marginal and joint frequency distribution, $\hat{f}[r]$ and $f[a,r]$, become increasingly leptokurtic. Figure \ref{fig:nf} shows the effect of the negative feedback through variations in $\delta$.

Even though the feedback from agents' actions to the outcome is one of the foundational blocks of reflexivity theory \citep{Soros2013}, negative feedback can be somehow questionable within financial markets. Indeed, asset bubble episodes are normally explained as positive feedback mechanisms that unfold through strategic complementarities. Despite Soros himself acknowledging negative feedback as the mechanism aligning individual beliefs with objective reality, in our model, it has a different implication, which concerns the liquidity effect.

If too many agents adopt the same trading strategy, they effectively reduce the overall profitability of that strategy, generating a crowding-out effect. Accordingly, this endogenous process tends to push the market outcome towards a conventional rate of return, which satisfies investors' behavior, thus explaining the observed peakedness in the distribution of rates of return.

\subsection{The Role of Expectations} \label{exp}
The perception of the social outcome determines agents’ actions through their estimated fundamentals $\mu$. Because their actions determine the outcome $r$, if individual expectations are self-fulfilling, that is to say, they are aligned with the central tendency of the outcome ($\bar{r}$), then $\mu=\bar{r}$. However, agents’ expectations about their fundamental valuation do not necessarily need to be correct. Accordingly, when $\mu\neq\bar{r}$, expectations are unfulfilled, and this provides an incentive for agents to reassess their estimate of the fundamentals through some market-based punishment for acting in a contrarian fashion with the market. On the contrary, when expectations are fulfilled ($\mu=\bar{r}$), there are no market-based incentives for agents to change their estimate of $\mu$ \citep{Foley2020}.

An important implication of the degree of expectation fulfillment is that whenever agents have mistaken beliefs, the resulting statistical equilibrium distribution of social outcomes becomes asymmetric. Our model captures skewness by the relation between $\mu$ and $\alpha$. When $\mu=\alpha$, meaning that the fundamental valuation coincides with the level of returns towards which the buying and selling actions push the market outcomes, the resulting marginal distribution is symmetric. On the contrary, when $\mu\neq\alpha$, then we can either have positive ($\mu<\alpha$) or negative ($\mu>\alpha$) skewness. Accordingly, we can distinguish between three different cases:
\begin{itemize}
    \item $\mu>\alpha \Rightarrow \, \bar{r}>\alpha$
    \item $\mu=\alpha \Rightarrow \, \bar{r}=\alpha$
    \item $\mu<\alpha \Rightarrow \, \bar{r}<\alpha$
\end{itemize}

\noindent We can prove this by considering Equation \ref{eq:log_odds}:
\begin{equation*}
    \log{\left[\frac{f[buy|r]}{f[sell|r]}\right]}=\frac{r-\mu}{T}
\end{equation*}

\noindent Since, in statistical equilibrium, for a given $\overline{r}$, $f[buy|\overline{r}]=f[sell|\overline{r}]$, then we have:
\begin{equation*}
    \log{[1]}=0=\frac{\overline{r}-\mu}{T} \Rightarrow \overline{r}=\mu
\end{equation*}

\noindent With this equilibrium condition in mind, we can derive the three above cases.

By comparing $\mu$ with $\alpha$ we are effectively analyzing the deviations of a subjective valuation from a ``conventional'' rate of return, resulting from the unintended consequences of investors' behavior. In particular, if $\mu=\alpha=\bar{r}$, then the center of gravitation around which market returns fluctuate is the same as the average rate of return. Therefore, our model allows us to quantify the deviation of agents' expectations about fundamentals from the actual location of the market as follows:
\begin{equation}
    \zeta=\mu-\alpha
\end{equation}

This implies that we can think of $\zeta$ as a measure of expectation fulfillment, which allows us to compare actual market outcomes with the Efficient Market Hypothesis (EMH) proposition \citep{Fama1965b}. Indeed, since the EMH assumes rational (self-fulfilling) expectations, which means $\zeta=0$, we can detect the presence of ``bubbles" in the sense of divergences of actual market values from fundamentals and thus provide an assessment of the EMH.

\begin{figure}[H]
	\centering
	\caption{Marginal, conditional, and joint frequency distributions for fulfilled and unfulfilled expectations.}
	\begin{subfigure}{.7\textwidth}
		\includegraphics[width=.9\textwidth]{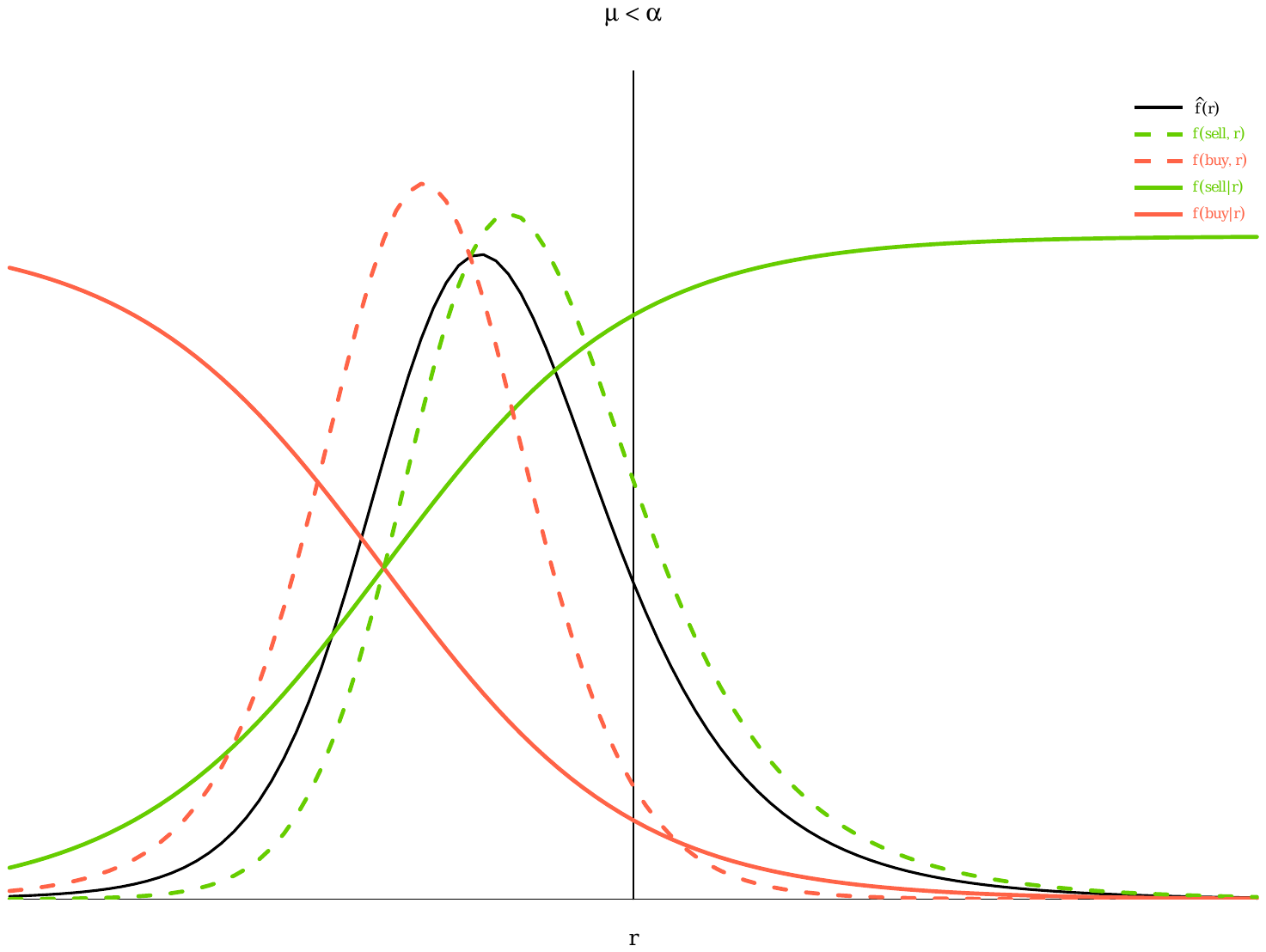}
		\caption{\scriptsize Positive skewness: $\zeta<0$ ($\mu=-2,\,T=1,\,\alpha=0$).}
	\end{subfigure}
	\begin{subfigure}{.7\textwidth}
		\includegraphics[width=.9\textwidth]{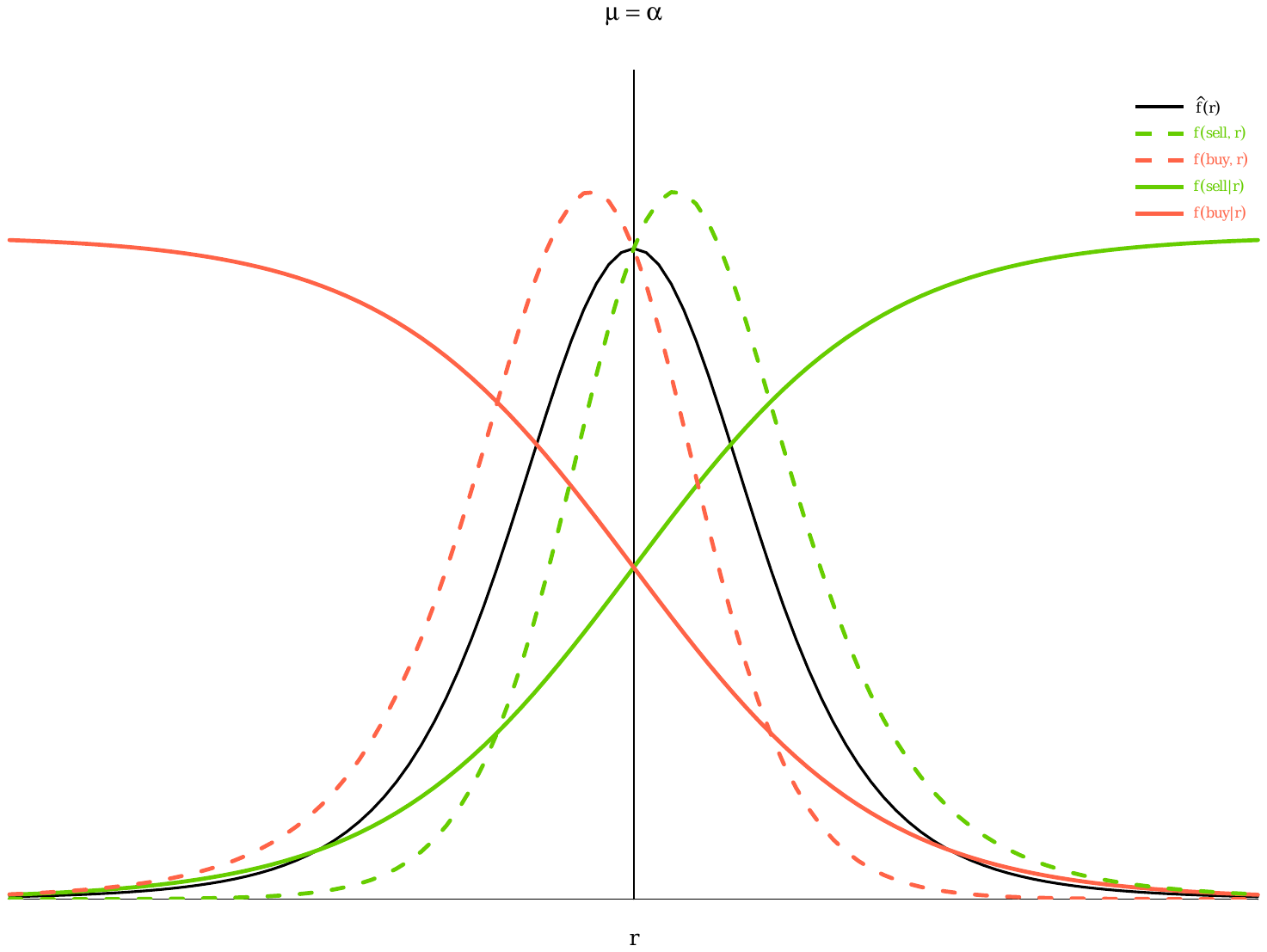}
		\caption{\scriptsize Symmetry: $\zeta=0$ ($\mu=\alpha=0,\,T=1$).}
	\end{subfigure}
	\begin{subfigure}{.7\textwidth}
		\includegraphics[width=.9\textwidth]{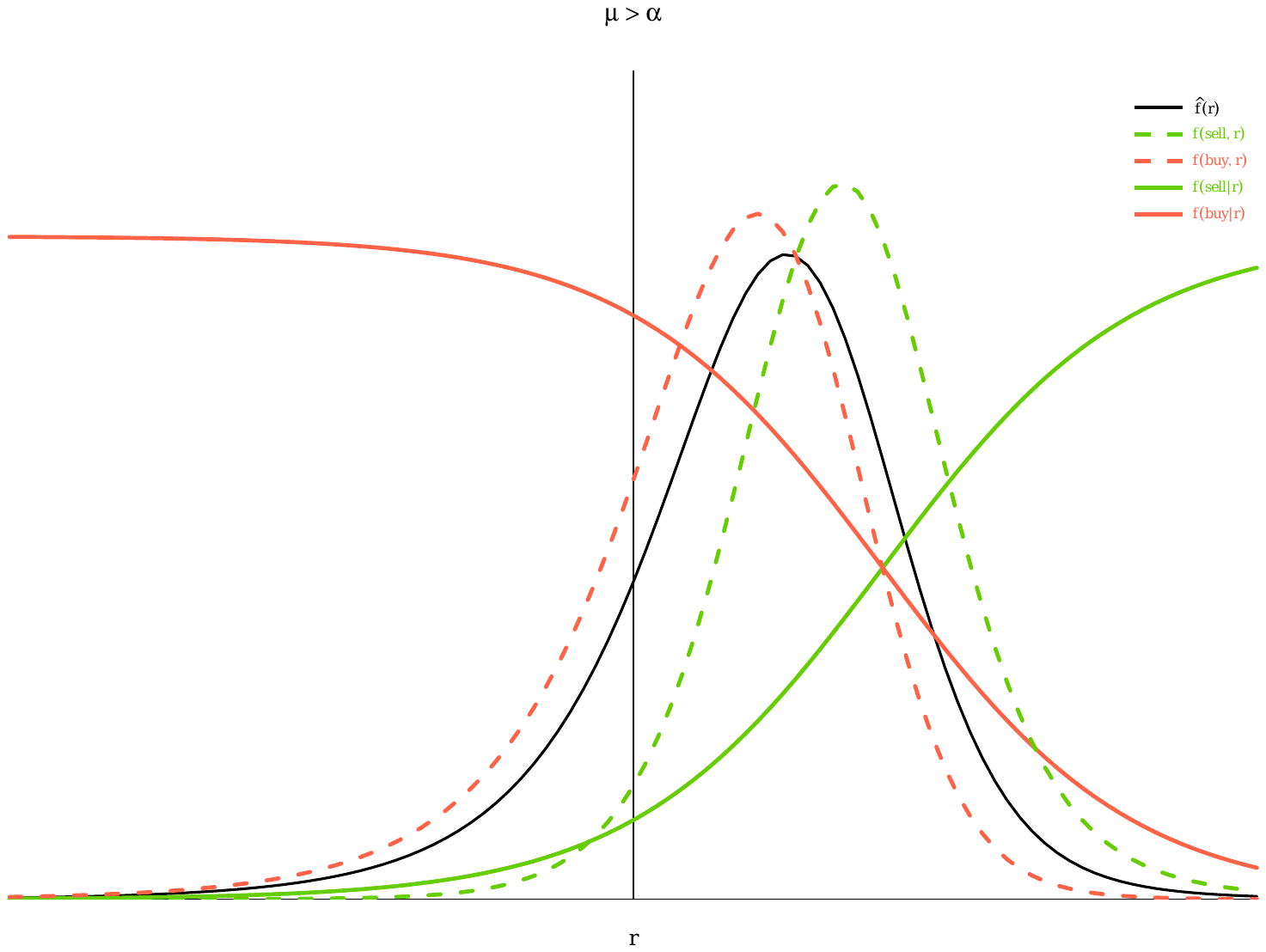}
		\caption{\scriptsize Negative skewness: $\zeta>0$ ($\mu=2,\,T=1,\,\alpha=0$).}
	\end{subfigure}
	\label{fig:exp}
\end{figure}

Figure \ref{fig:exp} shows the marginal, joint, and conditional probability distributions for fulfilled and unfulfilled expectations. As we can see, positive skewness can be interpreted as a ``market punishment for buyers”, meaning that they are buying at a higher price increase as compared to the conventional market valuation. On the contrary, negative skewness denotes a ``market punishment for sellers”, whose willingness to sell declines as they expect a rate of return above the average. An important implication of this process is that it imposes costly market punishment mechanisms associated with the correction of expectations, which leads to inertia in the adjustment of the system (as opposed to the rational expectations hypothesis, which assumes that such adjustments are instantaneous and costless).

\section{Discussion of Results} \label{res}
In what follows, we fit the QRSE model to the cross-sectional distribution of returns over four trading months rolling windows (80 days for the S\&P 500 companies data, 120 days for cryptocurrencies). Table \ref{tab:sum_res} summarizes the main results of the model over bull and bear markets.

\begin{table}[H]
   \centering
   \caption{Summary of Model Estimates (\%/day).}
   \scalebox{.7}{
    \begin{tabular}{cccccccccc}
	\hline \\[-1.8ex]
    & & ID & $\mu$ & $T$ & $\alpha$ & $S$ & $\bar{r}$ & $\zeta$ & $\delta$\\ [.5ex]
    \hline \hline \\[-1ex]
	\multirow{4}{4em}{S\&P 500} & Bull & 0.0099 & 0.3063 & 1.0574 & -0.1592 & 1.6907 & 0.0407 & 0.4655 & 0.9297\\[1ex]
    & & (0.0029) & (0.2319) & (0.3929) & (0.2988) & (0.6877) & (0.0979) & (0.5149) & (0.4366)\\[1ex] \\[-1ex]
    & Bear & 0.0092 & 0.1709 & 1.3817 & 0.0283 & 1.9603 & 0.0891 & 0.1427 & 1.0832\\[1ex]
    & & (0.0009) & (0.2602) & (0.3826) & (0.2714) & (0.4841) & (0.1284) & (0.4775) & (0.2888)\\ \hline \\[-1ex]
    \multirow{4}{4em}{Crypto} & Bull & 0.1391 & -0.0738 & 2.1997 & 0.0131 & 7.0157 & 0.1930 & -0.0869 & 5.0012\\[1ex]
    & & (0.0261) & (0.7184) & (1.5311) & (1.1933) & (1.5406) & (0.7227) & (1.8040) & (1.9376)\\[1ex] \\[-1ex]
    & Bear & 0.1322 & 0.2234 & 2.3296 & -0.5803 & 5.8279 & -0.2274 & 0.8037 & 3.6391\\[1ex]
    & & (0.0410) & (0.3879) & (0.6872) & (0.8071) & (0.8856) & (0.4195) & (1.1102) & (0.6821)\\ \hline
    \end{tabular}}
    \\ \medskip \small Average values of parameters and associated standard deviations (in parenthesis).
    \label{tab:sum_res}
\end{table}

As we can see, the goodness of fit, as captured by the Soofi ID, is better for the S\&P 500 data than cryptocurrencies. This is not surprising, given the bigger sample size. The same reasoning applies to the standard deviation of the estimates. We now focus our attention on the major parameters of the model: $\zeta$ (which captures the deviation between $\mu$ and $\alpha$), $T$, and $S$ (the inverse of the Lagrangian multiplier of the feedback constraint)\footnote{See Appendix \hyperref[app]{D} for further details.}, and $\delta$.

\begin{figure}[H]
	\centering
	\caption{Cross-sectional Estimates of $\zeta$.}
	\begin{subfigure}{.57\textwidth}
		\includegraphics[width=\textwidth]{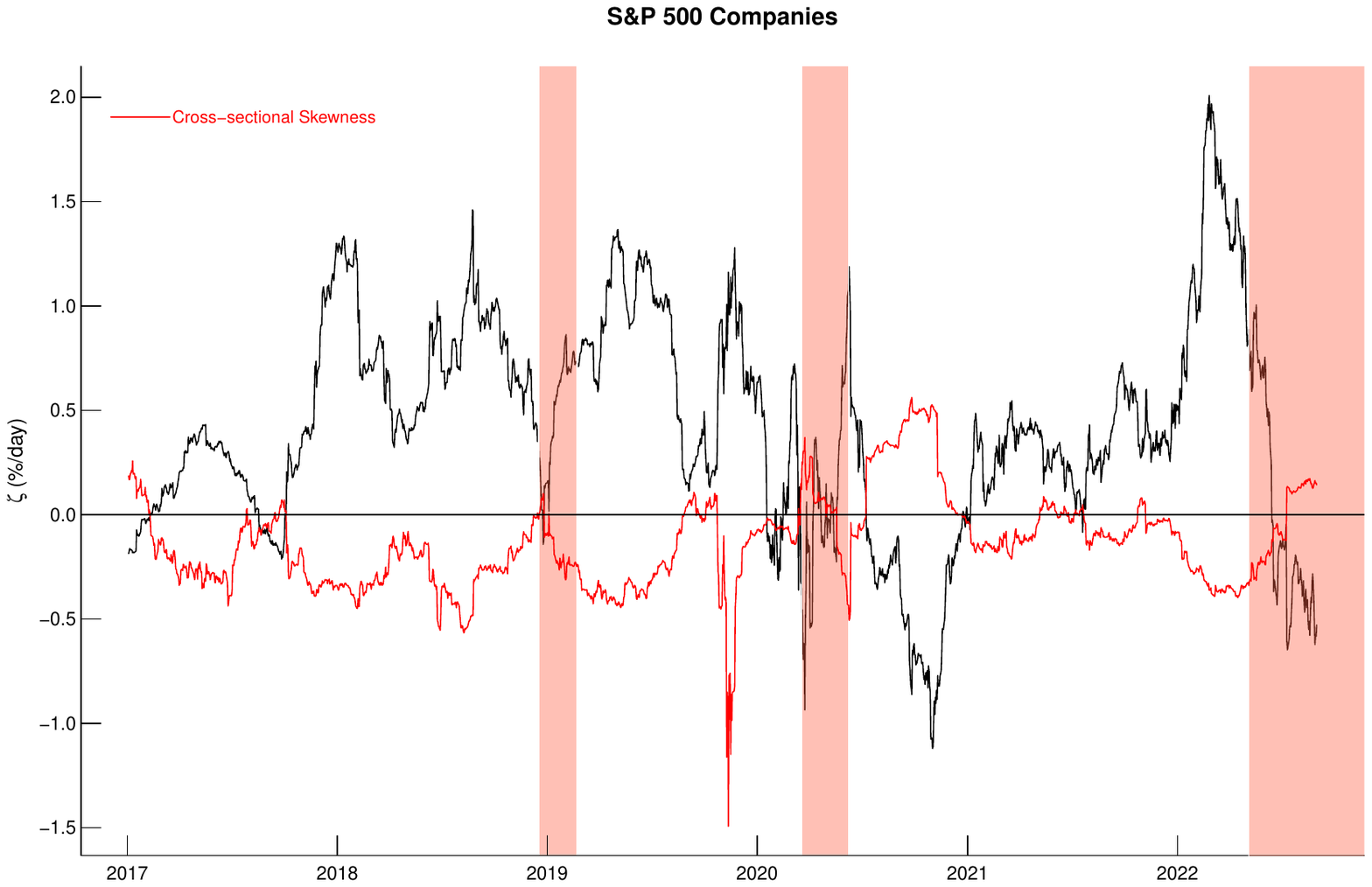}
		\caption{S\&P 500 Companies.}
	\end{subfigure}
	\begin{subfigure}{.57\textwidth}
		\includegraphics[width=\textwidth]{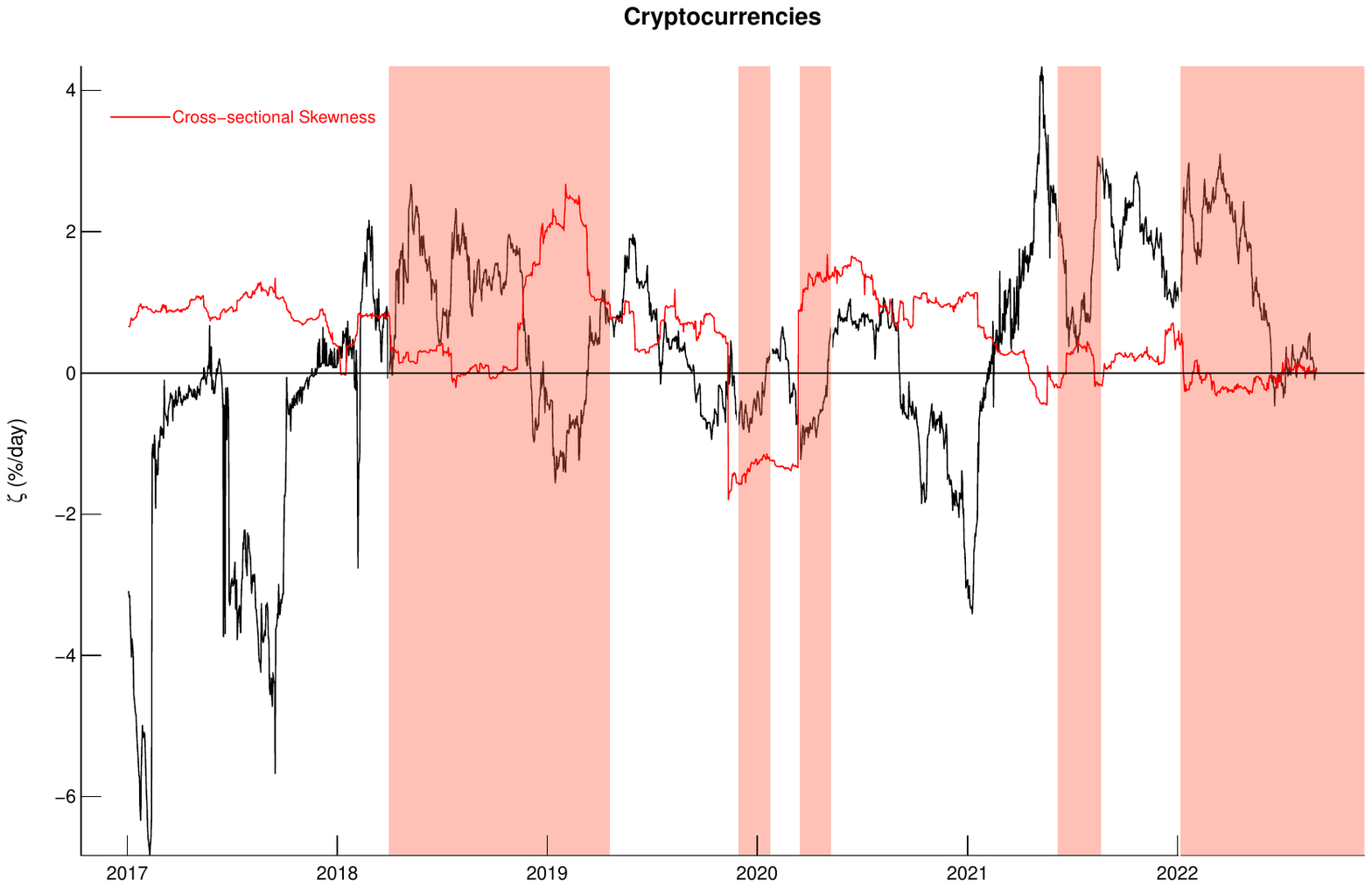}
		\caption{Cryptocurrencies.}
	\end{subfigure}
    \\ \footnotesize Red shaded denote bear markets. Estimates are shown over four-month rolling windows.
	\label{fig:zeta_est}
\end{figure}

Figure \ref{fig:zeta_est} shows the estimates of $\zeta$ for both the S\&P 500 and the cryptocurrencies' cross-sectional distributions. Recall that $\zeta$ is inversely related to the empirical cross-sectional skewness. In both cases, we can observe a decreasing trend in the degree of unfulfilled expectations preceding bear markets followed by a subsequent increase. This results from the correction of expectations of excessively buoyant investors, who buy at a higher rate of return ($\mu$) compared to the conventional valuation of the market ($\alpha$), in light of changes in realized outcomes. However, this trend is more irregular for cryptocurrencies, which denotes prolonged periods of bear markets. In this case, we witness an increase in the range of variations of $\zeta$ and faster adjustments to market changes. We attribute this to the distinctive nature of the cryptocurrency market, which trades twenty-four hours a day over the entire week.

We can acknowledge two other distinguishing features of the markets. First, the cross-sectional distribution of the S\&P 500 companies is mostly negatively skewed, as opposed to the bulk of the cryptocurrencies' distribution, which lies in the negative quadrant. This features the structural differences between the two markets. Second, there is a higher degree of mean reversion towards zero in the cryptocurrency market than in the stock market. This implies, according to our analysis, a more efficient market. Indeed, the average value of $\zeta$ for cryptocurrency over the entire sample is 0.27\%/day, as opposed to 0.42\%/day for the S\&P 500. This is due to a more intense arbitrage activity across cryptocurrency exchanges, on the one hand, and the associated lack of regulatory measures, on the other \citep{MakarovSchoar2020}.
%In this respect, we should take into consideration the magnitude of the average daily trading volume in the cryptocurrency market, which in our dataset outpaces the stock market by a factor of 180.

\begin{figure}[H]
	\centering
	\caption{Trading Frequencies ($f[a]$) and Degree of Expectation Fulfillment ($\zeta$).}
	\begin{subfigure}{.45\textwidth}
		\includegraphics[width=\textwidth]{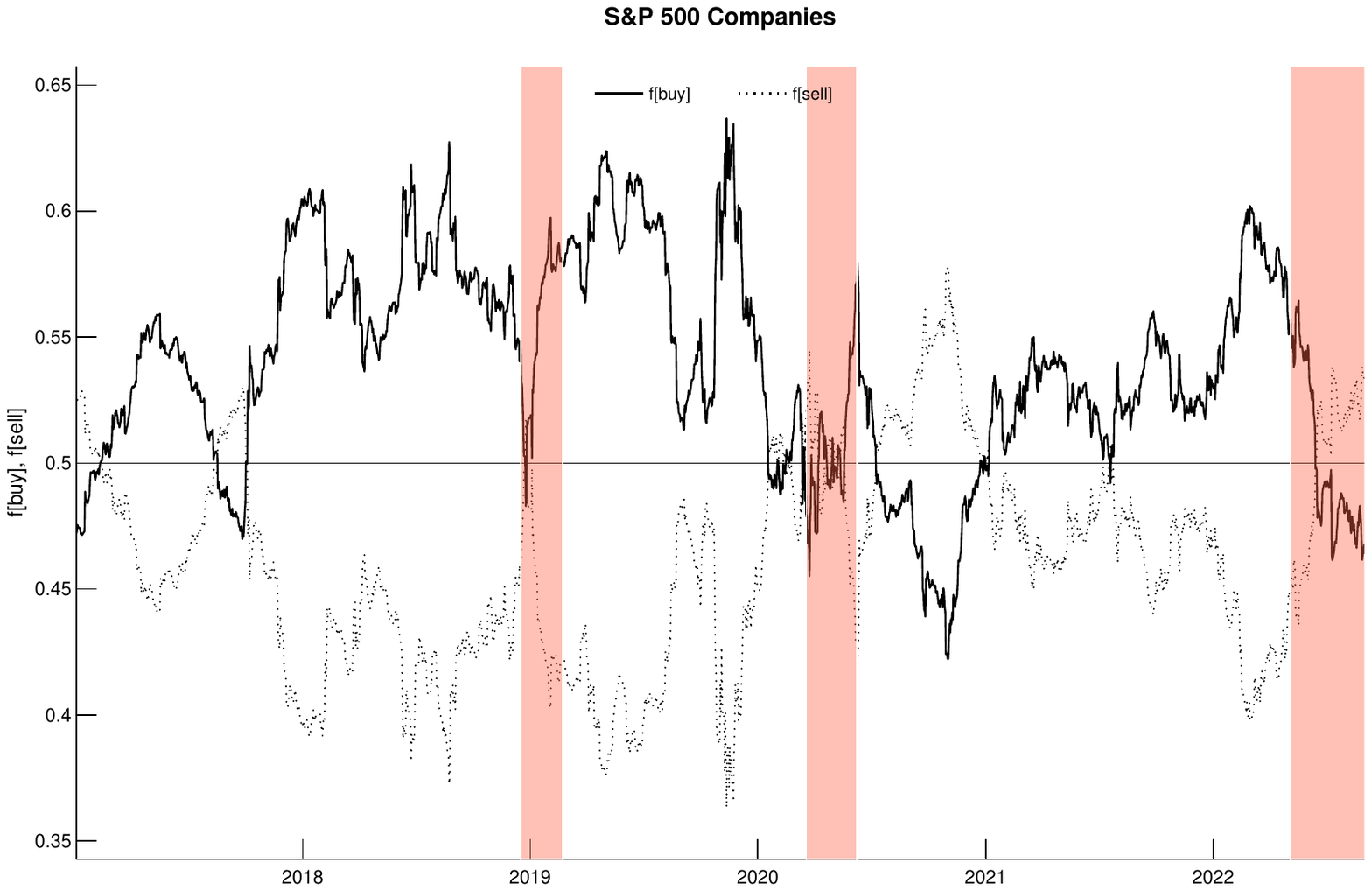}
		\caption{S\&P 500 Companies: $f[a]$.}
	\end{subfigure}
    \begin{subfigure}{.45\textwidth}
		\includegraphics[width=\textwidth]{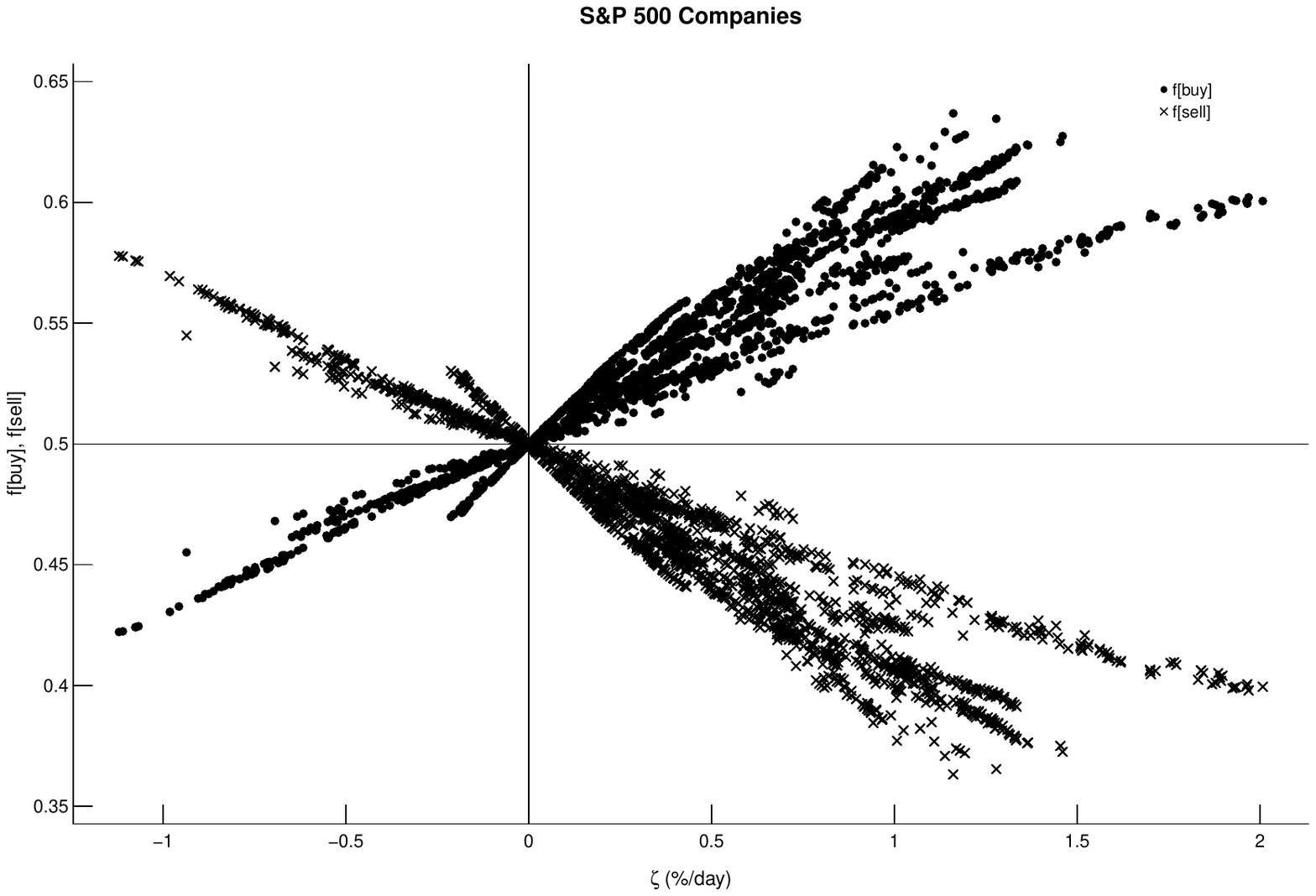}
		\caption{S\&P 500 Companies: $f[a]$, $\zeta$.}
	\end{subfigure}
    \begin{subfigure}{.45\textwidth}
		\includegraphics[width=\textwidth]{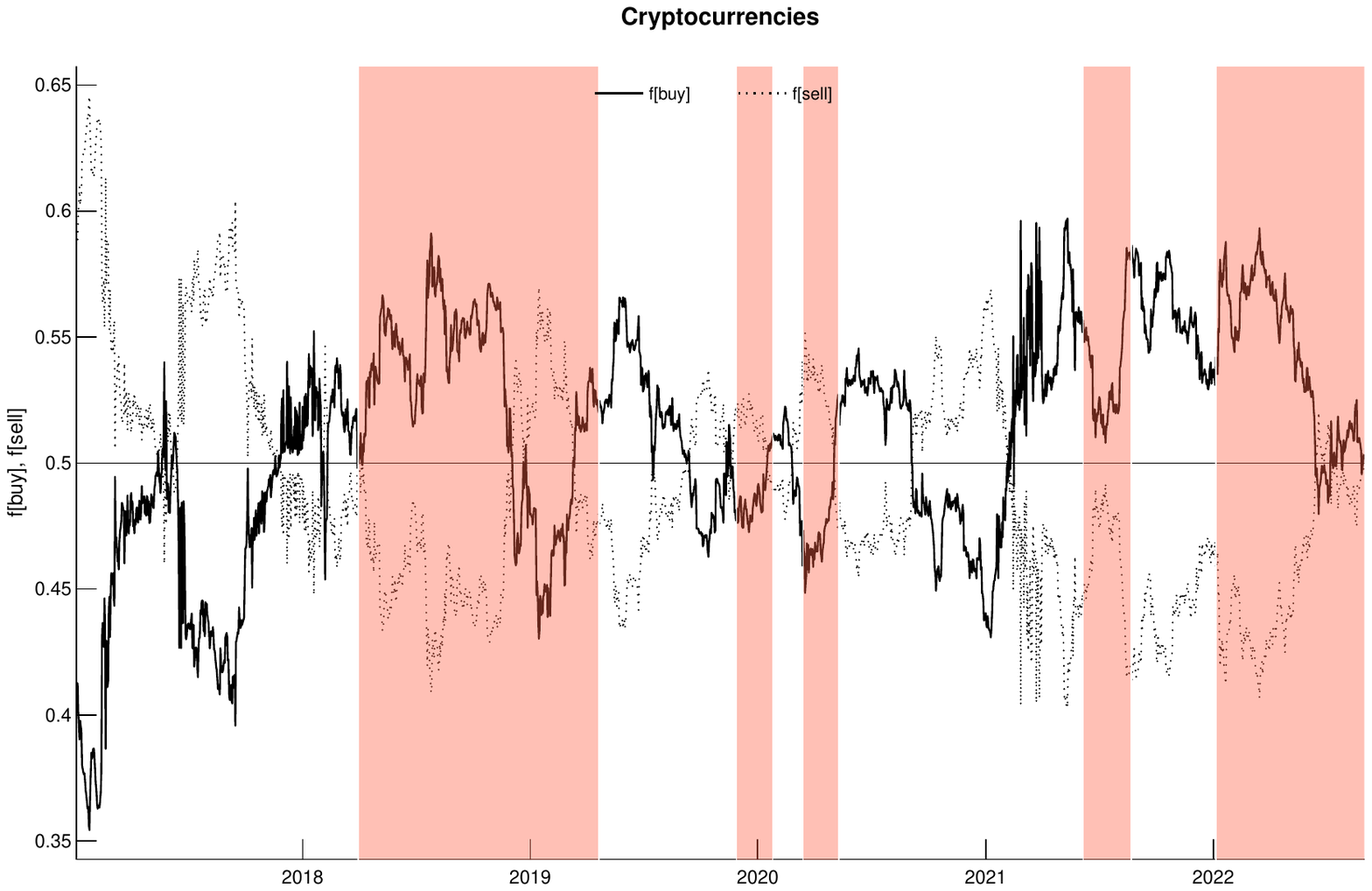}
		\caption{Cryptocurrencies: $f[a]$.}
	\end{subfigure}
	\begin{subfigure}{.45\textwidth}
		\includegraphics[width=\textwidth]{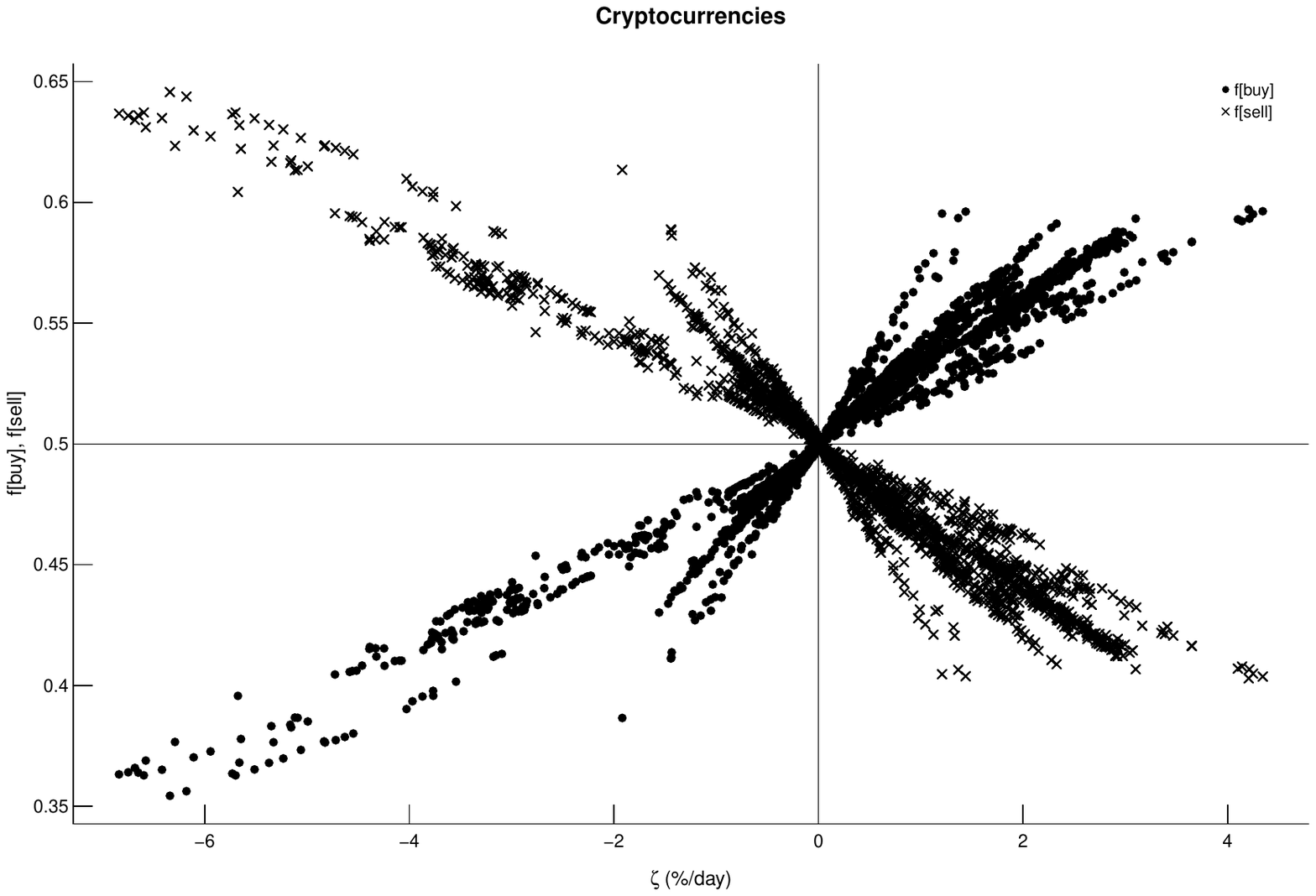}
		\caption{Cryptocurrencies: $f[a]$, $\zeta$.}
	\end{subfigure}
    \\ \footnotesize Red shaded areas denote bear markets.
	\label{fig:trad_freq}
\end{figure}

An interesting feature of the QRSE model is that it allows us to retrieve the trading frequencies, $f[buy]$ and $f[sell]$, by integrating the joint frequency distribution $f[a,r]$ over the rate or return $r$ as follows:
\begin{align}
\begin{split}
f[a] & = \int f[a,r] \, dr\\
& = \int f[a|r] f[r] \, dr
\end{split}
\end{align}
Figure \ref{fig:trad_freq} shows the implied trading frequency retrieved by the model and the relationship to the degree of expectation fulfillment for the cross-sectional distributions of the S\&P 500 companies, top panes, and the set of cryptocurrencies, bottom panes, respectively.\footnote{The symmetry between the buying and selling action is due to the property of the conditional frequency distribution, $f[a|r]$, which is symmetric around $\mu$.}

In terms of trading frequencies (top and bottom left panes), our results show a general decreasing trend in the buying frequencies before a bear market which is common to both asset classes. However, the intensity of trading is more remarkable in the cryptocurrency market, though bounded within a narrower range of values. This could be explained by a different degree of attentiveness of cryptocurrency investors to variations in price changes. Furthermore, the lack of a proper fundamental value for crypto assets, generates sudden reversals in trading activities.

If we relate the trading frequency to the degree of expectation fulfillment (top and bottom right panes), we acknowledge a direct relationship between the two variables. 
In detail, our results show a higher trading activity for positive values of $\zeta$ in the case of the S\&P 500 companies, whereas the opposite is true for the cryptocurrency market.
%This capture the idea that positive skewness in the cross-sectional distribution of rates of return ($\zeta<0$) is associated to a more intense selling activity, whereas negative skewness ($\zeta>0$) to increased buying frequencies. In this respect, 

\begin{figure}[H]
	\centering
	\caption{$T$ and $S$.}
	\begin{subfigure}{.48\textwidth}
		\includegraphics[width=\textwidth]{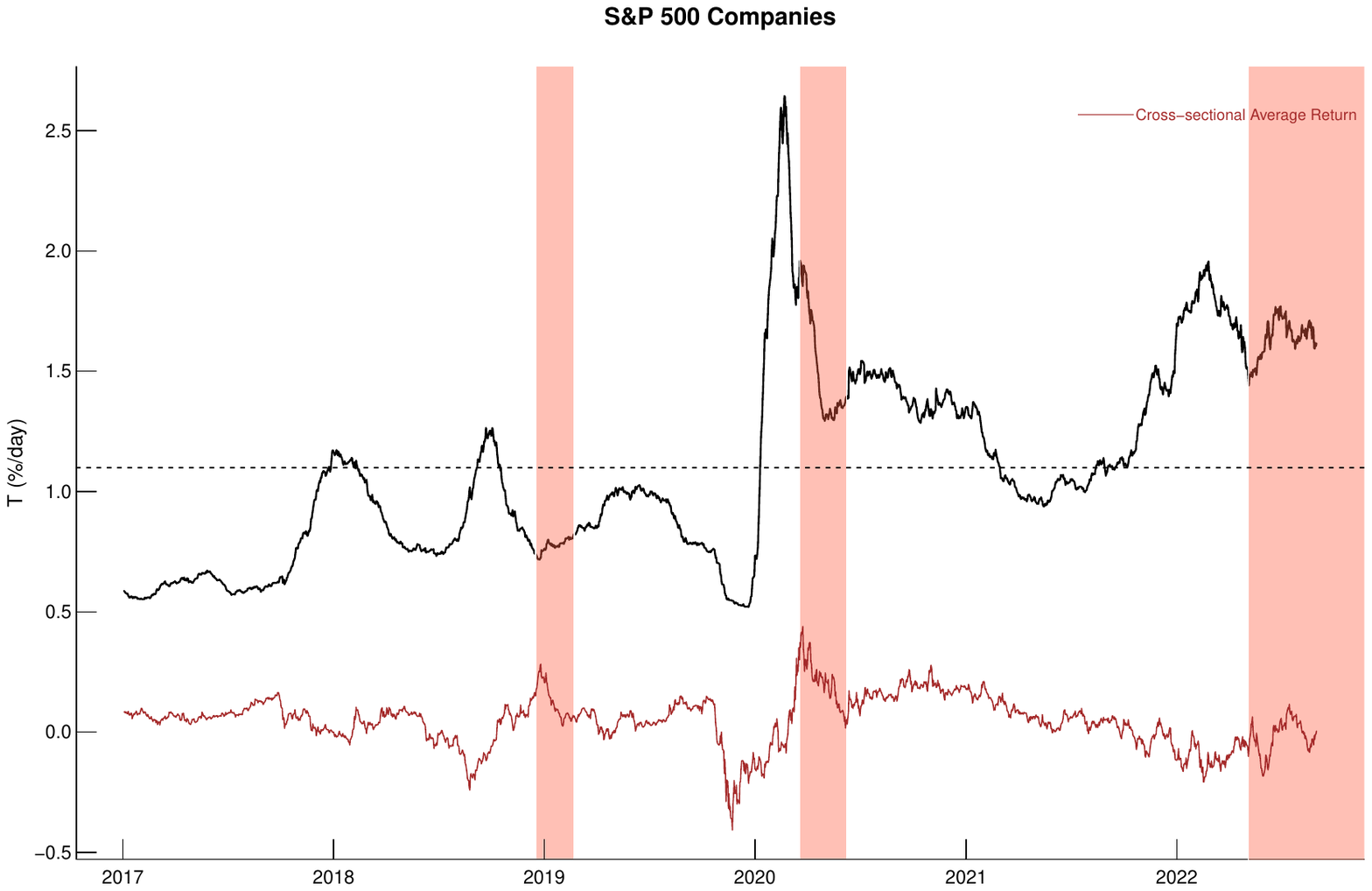}
		\caption{S\&P 500 Companies: $T$, $\bar{r}$.}
	\end{subfigure}
    \begin{subfigure}{.48\textwidth}
		\includegraphics[width=\textwidth]{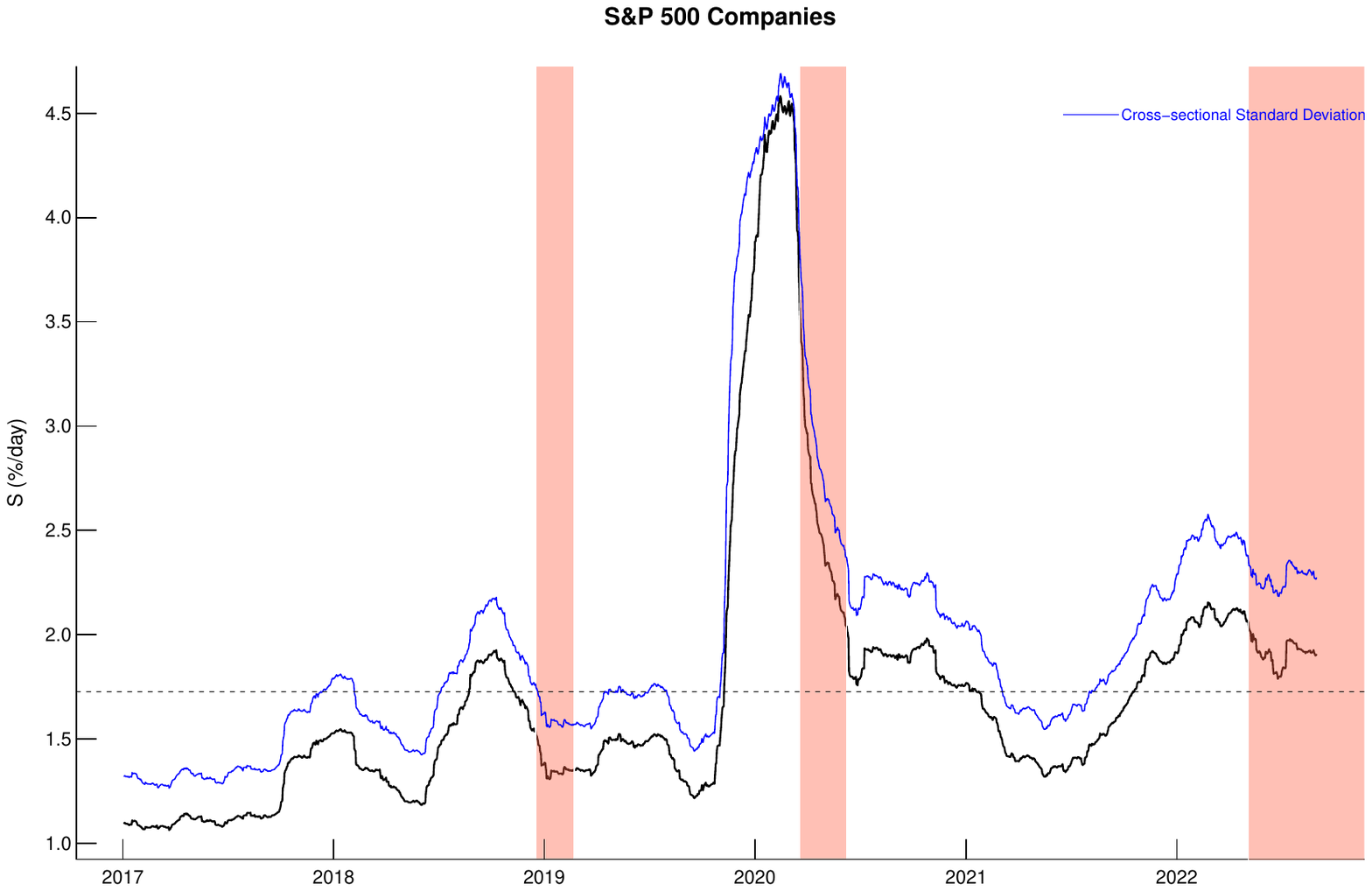}
		\caption{S\&P 500 Companies: $S$, $\sigma(r)$.}
	\end{subfigure}
    \begin{subfigure}{.48\textwidth}
		\includegraphics[width=\textwidth]{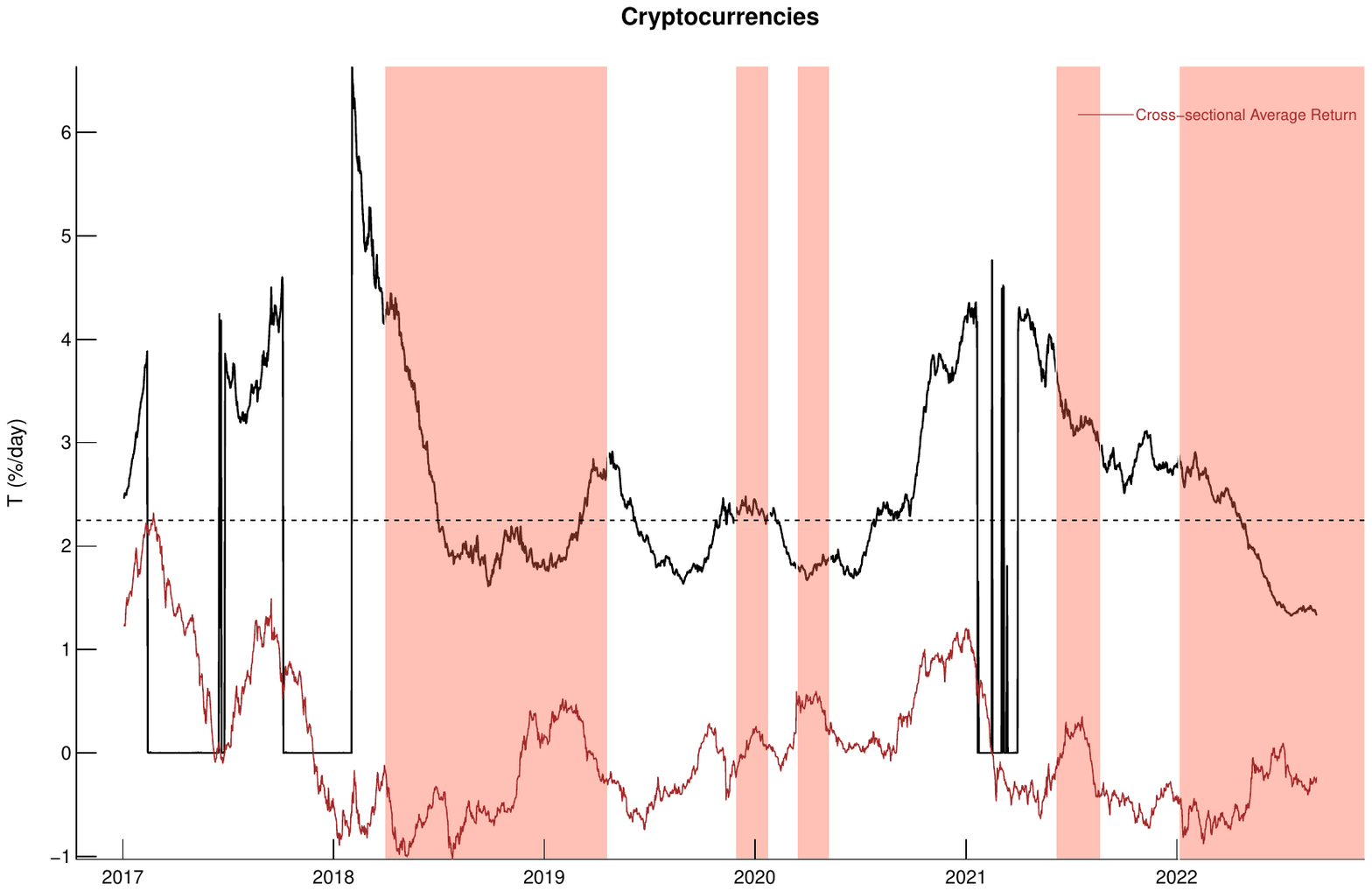}
		\caption{Cryptocurrencies: $T$, $\bar{r}$.}
	\end{subfigure}
	\begin{subfigure}{.48\textwidth}
		\includegraphics[width=\textwidth]{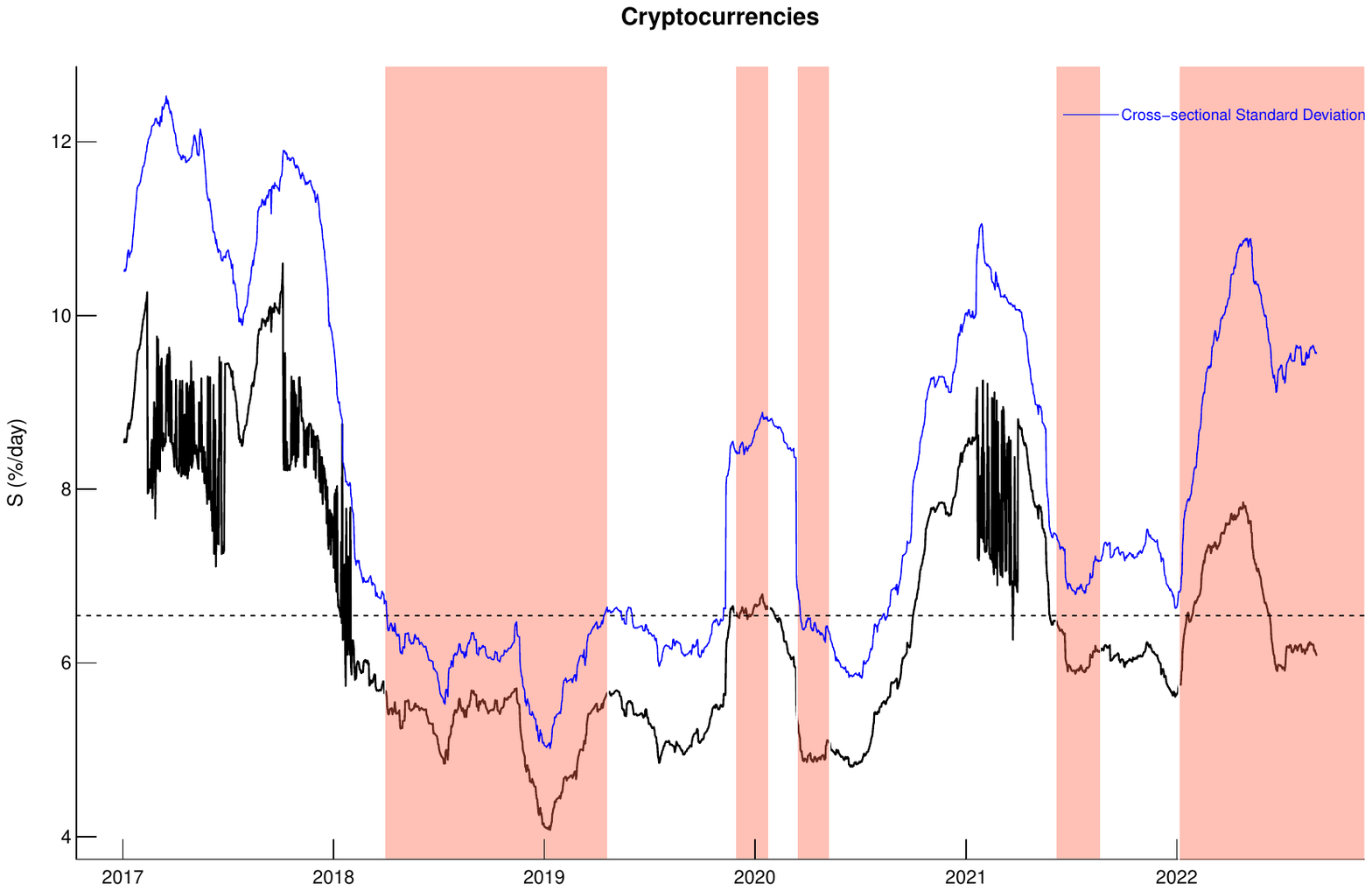}
		\caption{Cryptocurrencies: $S$, $\sigma(r)$.}
	\end{subfigure}
    \\ \footnotesize The dashed line shows the average value of the estimated parameter. Red-shaded areas denote bear markets.
	\label{fig:T_S}
\end{figure}

Figure \ref{fig:T_S} shows the cross-sectional estimates of $T$, left panes, and $S$, right panes, for the S\&P 500 companies and cryptocurrency data, respectively, along with the observed cross-sectional average ($\bar{r}$) and standard deviation ($\sigma(r)$) of the rate of return. $T$ is the microscopic parameter that captures the alertness of investors to variations in returns, whereas $S$, is a macroscopic parameter reflecting the aggregate fluctuation of the market.

Focusing on $T$, we can see that the range of variation is much wider for the S\&P 500 companies as opposed to cryptocurrencies. This captures the randomness in the behavior of cryptocurrency investors, who strongly overreact to variation in returns' differentials. Indeed, the average $T$ for the cryptocurrency cross-section is twice as big as the S\&P 500 companies. This implies, in terms of our analysis, that investors' actions are less predictable. We attribute this feature to the lack of a proper fundamental value for crypto assets, which is replaced by the ``madness of mobs". On the contrary, equity traders tend to rely on more solid conventions and anchor their expectations to stable patterns. We can verify this by comparing the time series of $T$ to the average cross-sectional return (brown line). For the S\&P 500 companies, the relation between the two is more stable, whereas it is hard to identify a specific trend in the cryptocurrency dataset. However, we can detect a common pattern across the two asset classes. Right before a bear market, $T$ suddenly declines and slowly reverts to its previous trend. This testifies to investors' panic right before a market collapse, which translates to an increased reaction to capital losses.

A similar argument can be made for $S$, on the right-hand pane of Figure \ref{fig:T_S}. For both datasets, we can see that this parameter properly tracks the observed cross-sectional volatility (blue line). Since the QRSE model provides a better fit for the S\&P 500 data, the estimated value of the parameter replicates almost exactly the empirical standard deviation.

\begin{figure}[H]
	\centering
	\caption{Relation between $\delta$ and Cross-sectional Kurtosis.}
	\begin{subfigure}{.6\textwidth}
		\includegraphics[width=\textwidth]{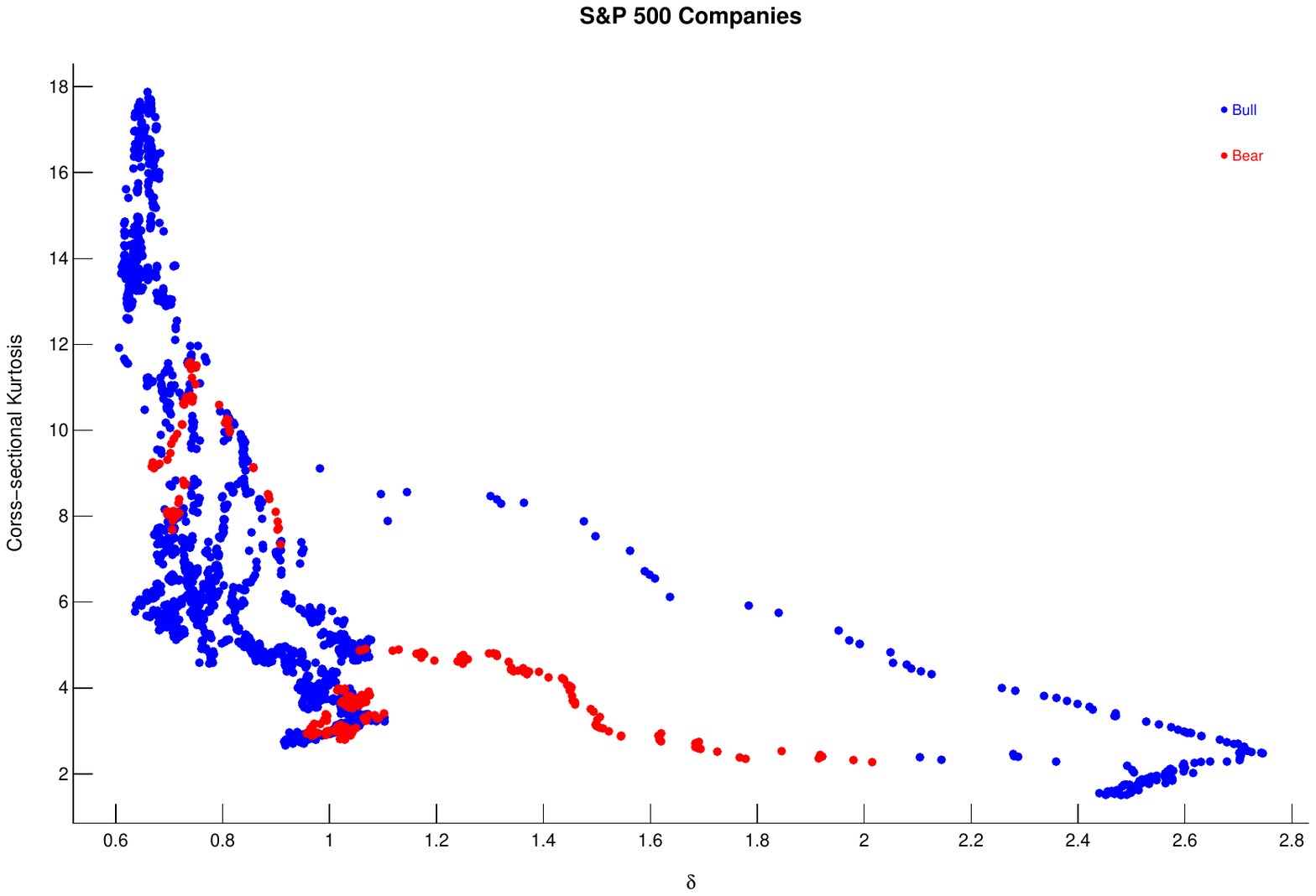}
		\caption{S\&P 500 Companies.}
	\end{subfigure}
    \hfill
	\begin{subfigure}{.6\textwidth}
		\includegraphics[width=\textwidth]{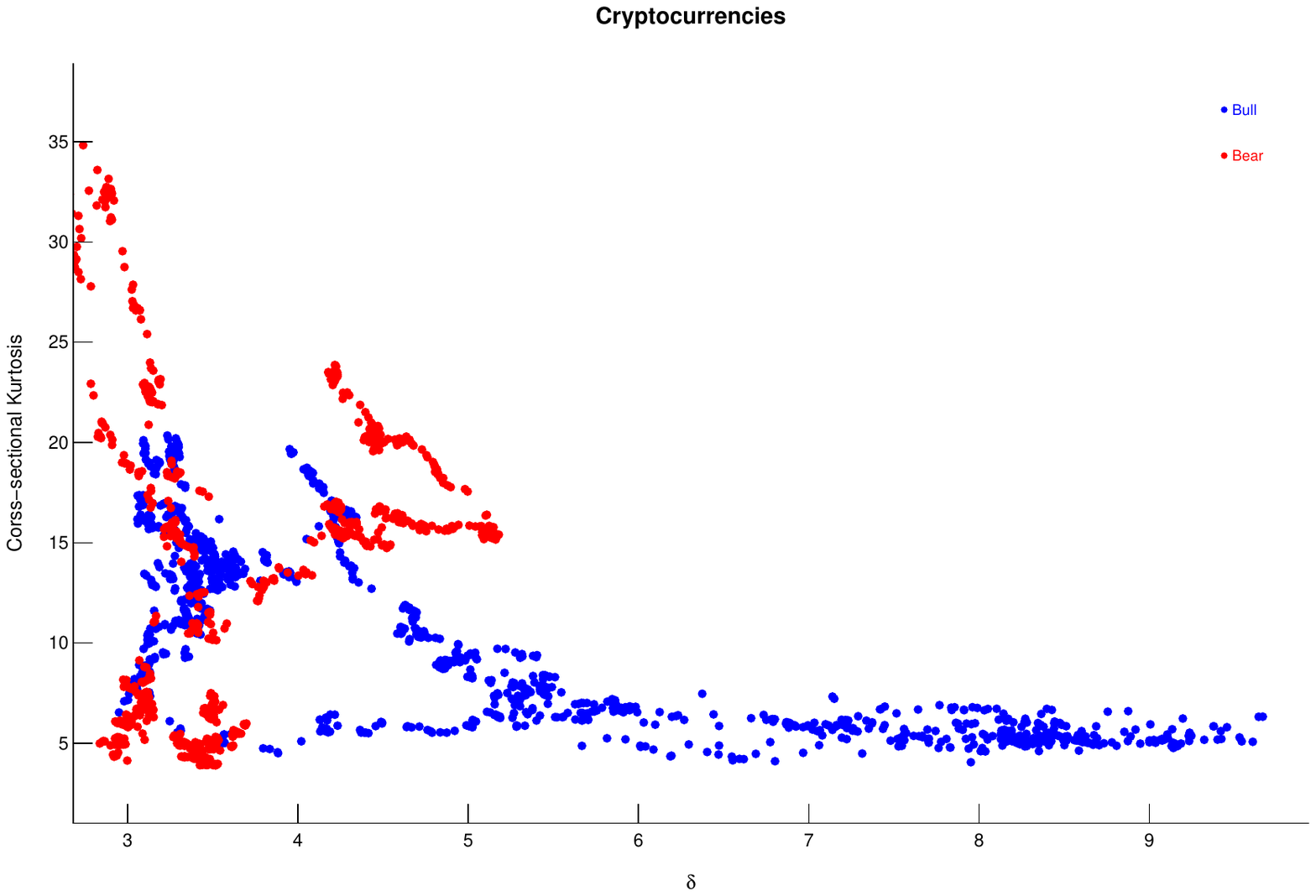}
		\caption{Cryptocurrencies.}
	\end{subfigure}
    \\ \footnotesize Red dots denote bear trends, blue dots bull trends.
	\label{fig:delta_est}
\end{figure}

We now turn to analyze $\delta$, which captures the effectiveness of the feedback constraint. As previously argued, the lower the $\delta$, the stronger the buying and selling actions in changing realized rates of return. In other words, $\delta$ gives an approximate measure of liquidity and the effectiveness of arbitrage in a market. Given that variations in $\delta$ are reflected in the kurtosis, Figure \ref{fig:delta_est} shows the relation between $\delta$ and the observed kurtosis of the cross-sectional distribution of the S\&P 500 companies, top pane, and the S\&P 500, bottom pane.

Both stock and cryptocurrency data confirm the negative relation between the two variables. However, we can appreciate the intensity of the negative feedback and the degree of competition in the stock market. This reflects differences in the structural features of the two markets, namely the degree of regulation and the role of arbitrage, that we discussed earlier. However, it is interesting to note how, for the S\&P 500 results, the majority of the estimates lie within a narrow range of values, at the bottom of the distribution, whereas for the cryptocurrency data, results are more spread out across the entire range of $\delta$.

\section{Conclusions} \label{concl}
The purpose of this paper is to shed light on the stochastic structure of the cryptocurrency and the stock market. By applying the notion of informational entropy, along with statistical equilibrium theory, we develop a model which allows us to study the cross-sectional distribution of daily returns of the companies listed in the S\&P 500 and a set of forty-eight cryptocurrencies.

On the one hand, our analysis confirms the differences in the structural features of the two markets. This is reflected both at the microscopic and the macroscopic level. Indeed, we can acknowledge the idiosyncratic features of equity and cryptocurrency traders in their alertness to variations in price changes. On the other, it quantifies the degree of unfulfilled expectations over time, thus providing a benchmark against the EMH. In this respect, our results show that the cryptocurrency market is more efficient than the stock market.

From a theoretical standpoint, a crucial implication of statistical equilibrium theory, as applied to financial markets, is that it explains how randomness in asset prices is not an accident of nature, but the unintended consequence of the action of market participants to profit from their information. If we relate this to the EMH, we can immediately acknowledge how it is possible to explain the same market outcome with completely different theoretical foundations. Whereas the EMH needs to resort to ``smart investors” to justify the stabilizing impact of speculation, the QRSE model tells us that the system can spontaneously converge towards a market convention, without implying agents' expectations to be fulfilled.

In addition to this, a constrained-entropy framework, combined with statistical equilibrium, allows us to infer how different types of frequency distributions generate regular patterns in financial time series, that the QRSE model can capture through interactions of its parameters. In this respect, understanding unfulfilled expectations might require the introduction of prior probabilities in the behavioral constraint \citep{EP2021}, asymmetries in the conditional action of buying and selling \citep{Black2018}, and the presence of multiple classes of market actors with heterogeneous beliefs \citep{SF2023}. These implementations will be the object of further research. 

\newpage
\section*{Appendix A. Cryptocurrency Dataset} \label{cryptos}
Our sample comprises forty-six cryptocurrencies:
\begin{itemize}
    \item 11 Proof-of-Work (PoW);
    \item 7 Non-Proof-of-Work (Non PoW);
    \item 28 additional ones (Extra).
\end{itemize}

\begin{table}[H]
   \centering
   \caption{List of Cryptocurrencies.}
   \scalebox{.67}{
    \begin{tabular}{cccc}
	\hline \\[-1.8ex]
    & Cryptocurrency & Average Daily Volume & Average Daily Market Cap\\ [.5ex]
    \hline  \hline \\[-1ex]
	\multirow{11}{4em}{PoW} & Bitcoin & 22,569,282,367.78 & 326,685,866,829.99\\
    & Dash & 343,331,104.28 & 1,492,295,764.57\\
    & Decred & 5,531,245.04 & 544,304,232.87\\
    & DigyBite & 18,936,982.50 & 312,951,299.64\\
    & Dogecoin & 870,544,228.14 & 6,844,697,073.31\\
    & Ethereum & 11,139,327,488.14 & 114,716,785,900.87\\
    & Ethereum Classic & 910,448,838.03 & 2,187,290,290.74\\
    & Litecoin & 2,055,950,930.04 & 5,796,974,190.01\\
    & Monero & 234,450,259.05 & 2,318,921,931.86\\ 
    & Vertcoin & 1,829,479.19 & 36,946,526.26\\
    & Zcash & 284,065,262.62 & 887,309,845.33\\ [1ex] \hline \\[-1ex]
    \multirow{7}{4em}{Non PoW} & Augur & 19,975,554.40 & 211,247,559.15\\
    & Lisk & 18,980,592.85 & 388,741,327.48\\
    & MadeSafeCoin & 1,050,644.33 & 134,416,818.31\\
    & NEM & 58,441,058.61 & 1,385,764,250.65\\
    & Stellar & 381,139,618.13 & 3,549,309,781.56\\
    & Waves & 115,061,196.54 & 682,943,285.98\\
    & XRP & 2,256,248,787.44 & 20,545,797,410.92\\ [1ex] \hline \\[-1ex]
    \multirow{29}{4em}{Extra} & Aeon & 62,217.50 & 10,997,964.15\\
    & Ardor & 9077770.48 & 167042607.80\\
    & BitShares & 15,525,158.95 & 195,030,928.48\\
    & Curecoin & 30,313.25 & 2,726,203.82\\
    & DigitalNote & 1,284,092.42 & 18,331,653.67\\
    & Einsteinium & 3,300,422.78 & 23,644,520.33\\
    & Expanse & 211,259.62 & 4,983,012.67\\
    & GameCredits & 779,839.16 & 37,581,204.06\\
    & Groestlcoin & 7,779,924.51 & 34,103,913.36\\
    & Memetic/PepeCoin & 272,306.74 & 1,781,065.27\\
    & MonaCoin & 4,705,228.86 & 100,184,067.73\\
    & MUNT & 100,505.42 & 14,416,092.92\\
    & Myriad & 88,776.17 & 4,484,131.87\\
    & Navcoin & 1,121,789.19 & 24,709,864.60\\
    & Neo & 329,434,192.41 & 1,670,353,730.08\\
    & Nexus & 1,032,469.22 & 43,424,690.22\\
    & Okcash & 468,809.03 & 4,802,979.06\\
    & Pinkcoin & 90,107.89 & 2,817,292.73\\
    & PIVX & 1,770,047.05 & 72,180,506.34\\
    & ReddCoin & 2,228,406.67 & 62,810,490.26\\
    & SaluS & 489,235.82 & 11,825,518.51\\
    & Signum & 39,649.36 & 11,699,222.57\\
    & Steem & 13,153,654.79 & 206,545,056.59\\
    & Steem Dollars & 6,481,324.82 & 22,062,800.04\\
    & Syscoin & 5,930,384.51 & 113,568,709.63\\
    & Validity & 772,715.76 & 9,863,786.05\\
    & Verge & 20,556,005.91 & 237,864,938.25\\
    & Viacoin & 886,564.72 & 14,481,368.84\\ [1ex] \hline
    \end{tabular}}
    \label{cryptolist}
\end{table}

\section*{Appendix B. The Maximum Entropy Principle} \label{MEP}
The Maximum Entropy Principle \citep{Jaynes2003} postulates the maximization of the entropy of an unknown and underdetermined distribution subject to constraints, expressing whatever information from observation or theory is relevant. Due to the strict concavity of the entropy function, as long as the constraints describing the available information represent a non-empty convex set in the space of distributions, the maximum entropy program will define some maximizing distribution that is a candidate statistical equilibrium of the model. The substantive interest of the resulting maximum entropy distribution depends on the information the constraints express.

Generally speaking, maximizing entropy maximizes the uncertainty of the system and gets the least informative state with no additional assumptions other than the existing knowledge of the researcher and observed data.  However, constrained maximum entropy can be used either as a method of ``rational inference” \citep{Golan2017}, where maximization of entropy resolves the residual uncertainty once the relevant information has been introduced via constraints, or as a ``statistical equilibrium theory”, which assumes that a system is observed in a state of equilibrium characterized by the decay of transients and is reaching a state of maximum entropy subject to constraints. Even though the formalism is the same for both methods, the interpretation of the results can be rather different. In our case, we deploy the maximum entropy principle as a statistical equilibrium theory.

The foundational concept in information theory is informational, or \cite{Shannon1948} entropy. In economic applications of information theory, entropy is most often described as a measure of uncertainty, but it is more helpful to think of entropy as the lack of predictability. Given a random variable $x\in X$, with probability distribution $f(x)$, the Shannon entropy, $\mathcal{H}[x]$ is defined as follows:
\begin{equation}
\mathcal{H}[x] = \mathbb{E}[-\log[f[x]] = - \sum_{x \in X} f[x]\log[f[x]]
\end{equation}

There are two reasons why we adopt this specific form of entropy. First, it is not a parametric measure, unlike \citet{Renyi1961} entropy and \citet{Tsallis1988} entropy, which are parametric and attach completely different weights to extremely rare and regular events. As suggested by \cite{Batra2020}, they might not be appropriate for analyzing financial data series. Second, Shannon entropy has been most successful in treating equilibrium systems, which is why it is adequate to deal with statistical equilibrium.

Let the rate of return on an asset be $r\in\mathbb{R}$ and $a\in\mathbb{A}$ the set of quantal actions of selling and buying stocks, $a=\{sell,buy\}$, respectively. Our objective is to determine the equilibrium joint distribution $f[a,r]$ with the marginal and conditional frequencies $f[r]$, $f[a]$, $f[a|r]$, $f[r|a]$ such that:
$$f[r]=\sum_{a}{f[a,r]} \, , \quad f[a]=\int_{r}f[a,r]\,dr \,;$$
$$f[a|r]=\frac{f[a,r]}{f[r]} \,\, \text{if} \, f[r]>0 \, , \quad f[r|a]=\frac{f[a,r]}{f[a]} \,\, \text{if} \, f[a]>0.$$

We write sums over outcomes, $r$, as integrals with the understanding that $r$ is treated as real-valued in theoretical applications. In empirical applications, however, measurements will inevitably be coarse-grained in a finite number of bins. We also omit the limits on integrals with the understanding that the sums are over the range of $r$.

If we were to maximize the entropy of the joint distribution $\mathcal{H}[f[a,r]]$, without the introduction of any constraint (except the normalization of the sum of the joint frequencies to unity), we would find that the entropy is maximized when the aggregate outcome $r$ is independent of the individual actions $a$. Since this result sheds no light on the process through which actions determine the outcome (it just returns the information already known through the observation of $f[r]$), we need to construct a theory of investors' behavior and how it impacts the social outcome by expressing it in terms of moment constraints.

\section*{Appendix C. Derivation of Investor's Behavior} \label{bc}
Let us assume that the typical agent’s response probability $f[a|r]$ depends on the payoff $u$ for choosing an action, which is the difference between the expected outcome variable $r$ and the agent’s expected average payoff or fundamental valuation of $r$, which we call $\mu$, such that we can write the payoff function as $u[a,r] = r-\mu$.

If an investor chooses a mixed strategy $f[a|r]: \mathbb{A} \times \mathbb{R} \rightarrow (0,1)$ to maximize the expected payoff $\sum_{a}{f[a|r]u[a,r]}$, then the informational entropy is:
\begin{equation}
\mathcal{H}[f[a|r]] = -\sum_{a} f[a|r]\log[f[a|r]]
\end{equation}

The entropy maximization program reads:
\begin{align}
\begin{split}
\underset{f[a|r]\geq 0}{\text{Max}}-&\sum_{a} f[a|r]\log[f[a|r]]\\
\text{subject to }& \sum_{a}f[a|r]=1\\
&\sum_{a} f[a|r]u[a,r]\ge U_{min}
\end{split}
\end{align}

Here, we are imposing a constraint on our uncertainty of $f[a|r]$ for a set of agents subject to the condition that individuals have a minimum expected payoff for acting (a sort of ``satisficing” behavior \emph{a $l\grave{a}$} \cite{Simon1955}). The associated Lagrangian has the following form:
\begin{equation}
\mathcal{L}= -\sum_{a} f[a|r]\log[f[a|r]] - \lambda \left(\sum_{a}f[a|r]-1 \right) + \beta \left( \sum_{a} f[a|r]u[a,r]- U_{min}\right)
\end{equation}

The solution to this program gives the maximum entropy distribution, which turns out to be the logit quantal response distribution (with $\beta=\frac{1}{T}$):\footnote{It is important to note that there are other ways to derive the above result through entropy maximization, and this is just one of them. \cite{Foley2020_art} thoroughly analyzes entropy-constrained behavior and its applications to economic theory.}
\begin{align} %\label{eq:lqr}
f[buy|r]&=\frac{1}{1+e^{\frac{u[a,r]}{T}}} &
f[sell|r]&=\frac{1}{1+e^{-\frac{u[a,r]}{T}}}
\end{align}

Our assumption about the agents' behavior implies that choice decisions are best described as a probabilistic phenomenon as opposed to the deterministic rational theory of choice, which assumes choices are always associated with probabilities equal to unity. In this sense, an interesting feature of the informational entropy-constrained model is that it gives meaning to the observed dispersion of behavior as the relative payoff of different actions, thus generating an ``entropy-constrained behavior". Interestingly, entropy-constrained behavior leads to the logit quantal response distribution without imposing any prior distributional assumption on the errors affecting the decision-making process.

\section*{Appendix D. Model Derivation and Inference} \label{app}
The maximum entropy problem that incorporates the behavioral and feedback constraints on the joint distributions reads as follows:
\begin{align}
\begin{split}
\underset{f[a,r]\geq 0}{\text{Max}}\mathcal{H}[a,r]=-& \int\sum_{a} f[a,r]\log[f[a,r]] \,dr\\
\text{ subject to } & \int\sum_{a}f[a,r] \,dr = 1\\
&\int \tanh \left[\frac{r-\mu}{2T}\right] \left(r-\alpha \right) f[r] \,dr \leq \delta
\end{split}
\end{align}

To solve this maximum entropy problem, it is convenient to write the joint entropy as the entropy of the marginal distribution plus the average entropy of the conditional distribution and solve for $f[r]$:
\begin{align}
\begin{split}
\mathcal{H}[a,r] =& -\mathcal{H}[r]+\int f[r]\mathcal{H}_{T,\mu}[r] \, dr\\
=&-\int f[r]\log[f[r]]dr+\int f[r]\sum_a f[a|r]\log[f[a|r]]  \, dr
\end{split}
\end{align}
where $\mathcal{H}_{T,\mu}[r]$ denotes the binary entropy function:
\begin{equation*}
\mathcal{H}_{T,\mu}[r] = -\sum_{a} f[a|r]\log[f[a|r]] = -\left( \frac{1}{1+e^{-\frac{r-\mu}{2T}}} \log \left[\frac{1}{1+e^{-\frac{r-\mu}{2T}}}\right] + \frac{1}{1+e^{\frac{r-\mu}{2T}}} \log\left[\frac{1}{1+e^{\frac{r-\mu}{2T}}}\right] \right)
\end{equation*}

The final maximum entropy program reads:
\begin{align}
\begin{split}
\underset{f[r]\geq 0}{\text{Max}} & -\mathcal{H}[r]+\int f[r]\mathcal{H}_{T,\mu}[r] \, dr\\
\text{     subject to } & \int f[r] \,dr = 1\\
&\int \tanh \left[\frac{r-\mu}{2T}\right] \left(r-\alpha \right) f[r] \,dr \leq \delta
\end{split}
\end{align}

This programming problem has the following associated Lagrangian:
\begin{align}
\begin{split}
\mathcal{L}[f[r],\lambda,\gamma] &= -\mathcal{H}[r] + \int f[r]\mathcal{H}_{T,\mu}[r]dr - \lambda \left(\int f[r]\,dr - 1\right) + \\
& \quad - \gamma\left(\int \tanh \left[\frac{r-\mu}{2T}\right] \left(r-\alpha\right) f[r] \,dr - \delta \right)
\end{split}
\end{align}

If we express $\gamma=\dfrac{1}{S}$, so we have an analog of the behavioral temperature for the feedback constraint, the first-order conditions for maximizing the entropy of the joint and conditional frequencies require:
\begin{equation}
\frac{\partial{\mathcal{L}}}{\partial{f[r]}} = -\log[f[r]]-1-\lambda+\mathcal{H}_{T,\mu}[r]-\tanh\left[\frac{r-\mu}{2T}\right]\left(\frac{r-\alpha}{S}\right) = 0
\end{equation}

The solution to this maximum entropy problem gives the most probable distribution of outcomes, that is, the marginal distribution $\hat{f}[r]$, that satisfies the constraints and has the following form:
\begin{equation} \label{eq:marg_r}
\hat{f}[r] = \frac{e^{\mathcal{H}_{T,\mu}[r] - \tanh\left[\frac{r-\mu}{2T}\right]\left(\frac{r-\alpha}{S}\right)}}{\int e^{\mathcal{H}_{T,\mu}[r] - \tanh\left[\frac{r-\mu}{2T}\right]\left(\frac{r-\alpha}{S}\right)}\, dr}
\end{equation}

The predicted marginal distribution $\hat{f}[r]$ from Equation (\ref{eq:marg_r}) is a Kernel to the maximum entropy program, and together with the parameters $\mu$, $T$, $\alpha$, and $S$ provides a multinomial distribution for the model. From a Bayesian perspective, the empirical marginal distribution, $\bar{f}[r]$ can be thought of as a sample of a multinomial model with frequencies $f[r]$ determined by Equation (\ref{eq:marg_r}). We use the Kullback-Leibler (KL) divergence as an approximation to the log posterior probability for the multinomial model since it allows us to make posterior inferences about the parameter estimates \cite[p.~15]{SF2017}.

The KL divergence measures the discrepancy between the empirical marginal frequencies, $\bar{f}[r]$, and the predicted marginal frequencies $\hat{f}[r;\mu,T,\alpha,S]$, inferred from the maximum entropy Kernel as:
\begin{equation}
D_{KL}[\hat{f}[r]||\bar{f}[r]] = \sum{\hat{f}[r] \log \left[\frac{\hat{f}[r]}{\bar{f}[r]}\right]}
\end{equation}

As a result, the KL divergence provides us with a tool to compare the observed marginal frequency distribution with the predicted marginal distribution. If $\hat{f}[r]=\bar{f}[r]$, the $D_{KL}$ becomes zero indicating that the two distributions are the same. Therefore, the smaller the KL divergence, the closer the observed distribution to the predicted one, and the better the fit is.

The set of parameters $\theta = \{\mu,T,\alpha,S\}$ are estimated jointly by minimizing the KL divergence.\footnote{The optimization algorithm for the KL minimization is the Broyden-Fletcher-Goldfarb-Shanno (BFGS) algorithm, with the following initial conditions for the parameters: $\mu=0$, $T=1$, $\alpha=0$, $S=1$.} To measure the closeness of the model fit, we use the information distinguishability criteria (ID) introduced by \cite{SR2002}. The ID measure shows approximately how much of the informational content of the observed frequencies is captured from the results of the maximum entropy program and is defined as follows:
\begin{equation}
ID[\hat{f}[r]:\bar{f}[r]] = 1 - e^{-D_{KL}[\hat{f}[r]||\bar{f}[r]]}
\end{equation}

\bibliographystyle{abbrvnat}
\bibliography{references}

\begin{thebibliography}{39}
\providecommand{\natexlab}[1]{#1}
\providecommand{\url}[1]{\texttt{#1}}
\expandafter\ifx\csname urlstyle\endcsname\relax
  \providecommand{\doi}[1]{doi: #1}\else
  \providecommand{\doi}{doi: \begingroup \urlstyle{rm}\Url}\fi

\bibitem[Batra and Taneja(2020)]{Batra2020}
L.~Batra and H.~Taneja.
\newblock Evaluating volatile stock markets using information theoretic
  measures.
\newblock \emph{Physica A: Statistical Mechanics and its Applications},
  537:\penalty0 122711, Jan. 2020.
\newblock \doi{10.1016/j.physa.2019.122711}.
\newblock URL \url{https://doi.org/10.1016/j.physa.2019.122711}.

\bibitem[Batra and Taneja(2022)]{Batra2022}
L.~Batra and H.~C. Taneja.
\newblock Comparison between information theoretic measures to assess financial
  markets.
\newblock \emph{{FinTech}}, 1\penalty0 (2):\penalty0 137--154, May 2022.
\newblock \doi{10.3390/fintech1020011}.
\newblock URL \url{https://doi.org/10.3390/fintech1020011}.

\bibitem[Bhambhwani et~al.(2023)Bhambhwani, Delikouras, and
  Korniotis]{Bhambhwani2023}
S.~M. Bhambhwani, S.~Delikouras, and G.~M. Korniotis.
\newblock Blockchain characteristics and cryptocurrency returns.
\newblock \emph{Journal of International Financial Markets, Institutions and
  Money}, 86:\penalty0 101788, 2023.
\newblock ISSN 1042-4431.
\newblock \doi{https://doi.org/10.1016/j.intfin.2023.101788}.
\newblock URL
  \url{https://www.sciencedirect.com/science/article/pii/S1042443123000562}.

\bibitem[Blackwell(2018)]{Black2018}
K.~Blackwell.
\newblock \emph{Entropy Constrained Behavior in Financial Markets. A Quantal
  Response Statistical Equilibrium Approach to Financial Modeling}.
\newblock PhD thesis, The New School, 2018.

\bibitem[Borri and Shakhnov(2022)]{Borri2022}
N.~Borri and K.~Shakhnov.
\newblock {The Cross-Section of Cryptocurrency Returns}.
\newblock \emph{The Review of Asset Pricing Studies}, 12\penalty0 (3):\penalty0
  667--705, 03 2022.
\newblock ISSN 2045-9920.
\newblock \doi{10.1093/rapstu/raac007}.
\newblock URL \url{https://doi.org/10.1093/rapstu/raac007}.

\bibitem[Brillouin(1951)]{Brillouin1951}
L.~Brillouin.
\newblock Maxwell{\textquotesingle}s demon cannot operate: Information and
  entropy. i.
\newblock \emph{Journal of Applied Physics}, 22\penalty0 (3):\penalty0
  334--337, Mar. 1951.
\newblock \doi{10.1063/1.1699951}.
\newblock URL \url{https://doi.org/10.1063/1.1699951}.

\bibitem[Brouty and Garcin(2022)]{Brouty2022a}
X.~Brouty and M.~Garcin.
\newblock {Maxwell's Demon and Information Theory in Market Efficiency: A
  Brillouin's Perspective}.
\newblock \emph{Physical Sciences Forum}, 5\penalty0 (1):\penalty0 23, Dec.
  2022.
\newblock \doi{10.3390/psf2022005023}.
\newblock URL \url{https://doi.org/10.3390/psf2022005023}.

\bibitem[Daw et~al.(2000)Daw, Finney, and Kennel]{Daw2000}
C.~S. Daw, C.~E.~A. Finney, and M.~B. Kennel.
\newblock Symbolic approach for measuring temporal
  {\textquotedblleft}irreversibility{\textquotedblright}.
\newblock \emph{Physical Review E}, 62\penalty0 (2):\penalty0 1912--1921, Aug.
  2000.
\newblock \doi{10.1103/physreve.62.1912}.
\newblock URL \url{https://doi.org/10.1103/physreve.62.1912}.

\bibitem[Delgado-Bonal(2019)]{DelgadoBonal2019}
A.~Delgado-Bonal.
\newblock Quantifying the randomness of the stock markets.
\newblock \emph{Scientific Reports}, 9\penalty0 (1), Sept. 2019.
\newblock \doi{10.1038/s41598-019-49320-9}.
\newblock URL \url{https://doi.org/10.1038/s41598-019-49320-9}.

\bibitem[Dickey and Fuller(1979)]{Dickey1979}
D.~A. Dickey and W.~A. Fuller.
\newblock Distribution of the estimators for autoregressive time series with a
  unit root.
\newblock \emph{Journal of the American Statistical Association}, 74\penalty0
  (366):\penalty0 427, June 1979.
\newblock \doi{10.2307/2286348}.
\newblock URL \url{https://doi.org/10.2307/2286348}.

\bibitem[Dickey and Fuller(1981)]{Dickey1981}
D.~A. Dickey and W.~A. Fuller.
\newblock Likelihood ratio statistics for autoregressive time series with a
  unit root.
\newblock \emph{Econometrica}, 49\penalty0 (4):\penalty0 1057, July 1981.
\newblock \doi{10.2307/1912517}.
\newblock URL \url{https://doi.org/10.2307/1912517}.

\bibitem[Evans and Prokopenko(2021)]{EP2021}
B.~P. Evans and M.~Prokopenko.
\newblock A maximum entropy model of bounded rational decision-making with
  prior beliefs and market feedback.
\newblock \emph{Entropy}, 23\penalty0 (6), 2021.
\newblock ISSN 1099-4300.
\newblock \doi{10.3390/e23060669}.
\newblock URL \url{https://www.mdpi.com/1099-4300/23/6/669}.

\bibitem[Fama(1965)]{Fama1965b}
E.~F. Fama.
\newblock Random walks in stock market prices.
\newblock \emph{Financial Analysts Journal}, 21\penalty0 (5):\penalty0 55--59,
  1965.
\newblock ISSN 0015198X.
\newblock URL \url{http://www.jstor.org/stable/4469865}.

\bibitem[Faynzilberg(1996)]{Faynzilberg1996}
P.~S. Faynzilberg.
\newblock Statistical mechanics of choice: {MaxEnt} estimation of population
  heterogeneity.
\newblock \emph{Annals of Operations Research}, 68\penalty0 (1):\penalty0
  161--180, Mar. 1996.
\newblock \doi{10.1007/bf02205453}.
\newblock URL \url{https://doi.org/10.1007/bf02205453}.

\bibitem[F.Donges et~al.(2013)F.Donges, Donner, and Kurths]{Donges2013}
J.~F.Donges, R.~V. Donner, and J.~Kurths.
\newblock Testing time series irreversibility using complex network methods.
\newblock \emph{{EPL} (Europhysics Letters)}, 102\penalty0 (1):\penalty0 10004,
  Apr. 2013.
\newblock \doi{10.1209/0295-5075/102/10004}.
\newblock URL \url{https://doi.org/10.1209/0295-5075/102/10004}.

\bibitem[Foley(2020{\natexlab{a}})]{Foley2020}
D.~K. Foley.
\newblock \emph{Unfulfilled Expectations: One Economist's History}.
\newblock Springer International Publishing, Cham, 2020{\natexlab{a}}.
\newblock ISBN 978-3-030-41357-6.
\newblock \doi{10.1007/978-3-030-41357-6_1}.
\newblock URL \url{https://doi.org/10.1007/978-3-030-41357-6_1}.

\bibitem[Foley(2020{\natexlab{b}})]{Foley2020_art}
D.~K. Foley.
\newblock Information theory and behaviors.
\newblock \emph{European Physical Journal Special Topics}, 229:\penalty0
  1591--1602, 2020{\natexlab{b}}.

\bibitem[Golan(2017)]{Golan2017}
A.~Golan.
\newblock \emph{{Foundations of Info-Metrics: Modeling, Inference, and
  Imperfect Information}}.
\newblock Oxford University Press, 12 2017.
\newblock ISBN 9780199349524.
\newblock \doi{10.1093/oso/9780199349524.001.0001}.
\newblock URL \url{https://doi.org/10.1093/oso/9780199349524.001.0001}.

\bibitem[Günther et~al.(2020)Günther, Fieberg, and Poddig]{Gunther2020}
S.~Günther, C.~Fieberg, and T.~Poddig.
\newblock The cross-section of cryptocurrency risk and return.
\newblock \emph{Vierteljahrshefte zur Wirtschaftsforschung}, 89:\penalty0
  7--28, 10 2020.
\newblock \doi{10.3790/vjh.89.4.7}.
\newblock URL \url{https://doi.org/10.3790/vjh.89.4.7}.

\bibitem[Jaynes(2003)]{Jaynes2003}
E.~T. Jaynes.
\newblock \emph{Probability theory: The logic of science}.
\newblock Cambridge University Press, 2003.

\bibitem[Kardar(2007)]{Kardar2007}
M.~Kardar.
\newblock \emph{Statistical Physics of Particles}.
\newblock Cambridge University Press, June 2007.
\newblock \doi{10.1017/cbo9780511815898}.
\newblock URL \url{https://doi.org/10.1017/cbo9780511815898}.

\bibitem[Kwiatkowski et~al.(1992)Kwiatkowski, Phillips, Schmidt, and
  Shin]{Kwiatkowski1992}
D.~Kwiatkowski, P.~C. Phillips, P.~Schmidt, and Y.~Shin.
\newblock Testing the null hypothesis of stationarity against the alternative
  of a unit root.
\newblock \emph{Journal of Econometrics}, 54\penalty0 (1-3):\penalty0 159--178,
  Oct. 1992.
\newblock \doi{10.1016/0304-4076(92)90104-y}.
\newblock URL \url{https://doi.org/10.1016/0304-4076(92)90104-y}.

\bibitem[Lo(2019)]{Lo2019}
A.~W. Lo.
\newblock \emph{Adaptive Markets: Financial Evolution at the Speed of Thought}.
\newblock Princeton University Press, 2019.
\newblock ISBN 9780691196800.
\newblock \doi{doi:10.1515/9780691196800}.
\newblock URL \url{https://doi.org/10.1515/9780691196800}.

\bibitem[Luque et~al.(2009)Luque, Lacasa, Ballesteros, and Luque]{Luque2009}
B.~Luque, L.~Lacasa, F.~Ballesteros, and J.~Luque.
\newblock Horizontal visibility graphs: Exact results for random time series.
\newblock \emph{Physical Review E}, 80\penalty0 (4), Oct. 2009.
\newblock \doi{10.1103/physreve.80.046103}.
\newblock URL \url{https://doi.org/10.1103/physreve.80.046103}.

\bibitem[Makarov and Schoar(2020)]{MakarovSchoar2020}
I.~Makarov and A.~Schoar.
\newblock Trading and arbitrage in cryptocurrency markets.
\newblock \emph{Journal of Financial Economics}, 135\penalty0 (2):\penalty0
  293--319, 2020.
\newblock ISSN 0304-405X.
\newblock \doi{https://doi.org/10.1016/j.jfineco.2019.07.001}.
\newblock URL
  \url{https://www.sciencedirect.com/science/article/pii/S0304405X19301746}.

\bibitem[Ramsey and Rothman(1996)]{Ramsey1996}
J.~B. Ramsey and P.~Rothman.
\newblock Time irreversibility and business cycle asymmetry.
\newblock \emph{Journal of Money, Credit and Banking}, 28\penalty0
  (1):\penalty0 1, Feb. 1996.
\newblock \doi{10.2307/2077963}.
\newblock URL \url{https://doi.org/10.2307/2077963}.

\bibitem[R{\'e}nyi(1961)]{Renyi1961}
A.~R{\'e}nyi.
\newblock On measures of entropy and information.
\newblock In \emph{Proceedings of the Fourth Berkeley Symposium on Mathematical
  Statistics and Probability, Volume 1: Contributions to the Theory of
  Statistics}, pages 547--561. University of California Press, 1961.

\bibitem[Scharfenaker and Foley(2017)]{SF2017}
E.~Scharfenaker and D.~Foley.
\newblock Quantal response statistical equilibrium in economic interactions:
  Theory and estimation.
\newblock \emph{Entropy}, 19\penalty0 (9):\penalty0 444, Aug. 2017.
\newblock \doi{10.3390/e19090444}.
\newblock URL \url{https://doi.org/10.3390/e19090444}.

\bibitem[Scharfenaker and Foley(2023)]{SF2023}
E.~Scharfenaker and D.~K. Foley.
\newblock The neutrality of money reconsidered: A statistical equilibrium model
  of the labor market.
\newblock 2023.
\newblock URL \url{https://EconPapers.repec.org/RePEc:uta:papers:2023_02}.

\bibitem[Shannon(1948)]{Shannon1948}
C.~E. Shannon.
\newblock A mathematical theory of communication.
\newblock \emph{Bell System Technical Journal}, 27\penalty0 (3):\penalty0
  379--423, July 1948.
\newblock \doi{10.1002/j.1538-7305.1948.tb01338.x}.
\newblock URL \url{https://doi.org/10.1002/j.1538-7305.1948.tb01338.x}.

\bibitem[Simon(1955)]{Simon1955}
H.~A. Simon.
\newblock A behavioral model of rational choice.
\newblock \emph{The Quarterly Journal of Economics}, 69\penalty0 (1):\penalty0
  99--118, 02 1955.
\newblock ISSN 0033-5533.
\newblock \doi{10.2307/1884852}.

\bibitem[Smith(1776)]{Smith1776}
A.~Smith.
\newblock \emph{An Inquiry into the Nature and Causes of the Wealth of
  Nations}.
\newblock McMaster University Archive for the History of Economic Thought,
  1776.

\bibitem[Soofi and Retzer(2002)]{SR2002}
E.~Soofi and J.~Retzer.
\newblock Information indices: unification and applications.
\newblock \emph{Journal of Econometrics}, 107\penalty0 (1):\penalty0 17--40,
  2002.
\newblock ISSN 0304-4076.
\newblock \doi{https://doi.org/10.1016/S0304-4076(01)00111-7}.
\newblock URL
  \url{https://www.sciencedirect.com/science/article/pii/S0304407601001117}.
\newblock Information and Entropy Econometrics.

\bibitem[Soros(2013)]{Soros2013}
G.~Soros.
\newblock Fallibility, reflexivity, and the human uncertainty principle.
\newblock \emph{Journal of Economic Methodology}, 20\penalty0 (4):\penalty0
  309--329, 2013.
\newblock \doi{10.1080/1350178X.2013.859415}.

\bibitem[Tran and Leirvik(2019)]{Tran2019}
V.~L. Tran and T.~Leirvik.
\newblock A simple but powerful measure of market efficiency.
\newblock \emph{Finance Research Letters}, 29:\penalty0 141--151, June 2019.
\newblock \doi{10.1016/j.frl.2019.03.004}.
\newblock URL \url{https://doi.org/10.1016/j.frl.2019.03.004}.

\bibitem[Tsallis(1988)]{Tsallis1988}
C.~Tsallis.
\newblock Possible generalization of boltzmann-gibbs statistics.
\newblock \emph{Journal of Statistical Physics}, 52\penalty0 (1-2):\penalty0
  479--487, July 1988.
\newblock \doi{10.1007/bf01016429}.
\newblock URL \url{https://doi.org/10.1007/bf01016429}.

\bibitem[Vințe et~al.(2021)Vințe, Ausloos, and Furtun{\u{a}}]{Vinte2021}
C.~Vințe, M.~Ausloos, and T.~F. Furtun{\u{a}}.
\newblock A volatility estimator of stock market indices based on the intrinsic
  entropy model.
\newblock \emph{Entropy}, 23\penalty0 (4):\penalty0 484, Apr. 2021.
\newblock \doi{10.3390/e23040484}.
\newblock URL \url{https://doi.org/10.3390/e23040484}.

\bibitem[Zanin and Papo(2021)]{Zanin2021}
M.~Zanin and D.~Papo.
\newblock Algorithmic approaches for assessing irreversibility in time series:
  Review and comparison.
\newblock \emph{Entropy}, 23\penalty0 (11):\penalty0 1474, Nov. 2021.
\newblock \doi{10.3390/e23111474}.
\newblock URL \url{https://doi.org/10.3390/e23111474}.

\bibitem[Zheng(2023)]{Zheng2023}
X.~Zheng.
\newblock Complex behavior of individuals and collectives in a social system:
  An introduction to exploratory computational experimental methodology based
  on multi-agent modeling.
\newblock \emph{Annals of Operations Research}, May 2023.
\newblock \doi{10.1007/s10479-023-05388-1}.
\newblock URL \url{https://doi.org/10.1007/s10479-023-05388-1}.

\end{thebibliography}

\end{document}